\documentclass[twocolumn,aps,prd,nofootinbib,superscriptaddress,showpacs]{revtex4-1}
\usepackage{amsmath,graphicx,bm,amsbsy,aas_macros,color,natbib,subfigure}
\pdfoutput=1

\newcommand{\BigO}[1]{\mathcal{O}(#1)}

\newcommand{\M}{\mathbf{M}}
\newcommand{\C}{\mathbf{C}}
\newcommand{\beq}{\begin{equation}}
\newcommand{\eeq}{\end{equation}}

\newcommand{\Eye}{\mathbf{I}}
\newcommand{\x}{\mathbf{x}}

\newcommand{\N}{\mathbf{N}}
\newcommand{\n}{\mathbf{n}}

\newcommand{\mean}{\boldsymbol\mu}
\newcommand{\base}{\mathbf{b}}

\newcommand{\A}{\mathbf{A}}
\newcommand{\p}{\mathbf{p}}

\newcommand{\Proj}{\mathbf{\Pi}}
\newcommand{\PSF}{\mathbf{P}} 
\newcommand{\K}{\mathbf{K}}
\newcommand{\y}{\mathbf{y}}
\newcommand{\J}{\mathbf{J}}
\newcommand{\D}{\mathbf{D}}

\newcommand{\T}{\mathbf{T}}

\newcommand{\Q}{\mathbf{Q}}
\newcommand{\E}{\mathbf{E}}
\newcommand{\trans}{\mathsf{T}}
\newcommand{\rhat}{\hat{\mathbf{r}}}

\begin{document} 

\title{Mapmaking for Precision 21\,cm Cosmology}

\author{Joshua S. Dillon}
\email{jsdillon@mit.edu}
\affiliation{Department of Physics, Massachusetts Institute of Technology, Cambridge, MA}
\affiliation{MIT Kavli Institute, Massachusetts Institute of Technology, Cambridge, MA}

\author{Max Tegmark}
\affiliation{Department of Physics, Massachusetts Institute of Technology, Cambridge, MA}
\affiliation{MIT Kavli Institute, Massachusetts Institute of Technology, Cambridge, MA}

\author{Adrian Liu}
\affiliation{Department of Astronomy, University of California Berkeley, Berkeley, CA}
\affiliation{Berkeley Center for Cosmological Physics, University of California Berkeley, Berkeley, CA}

\author{Aaron Ewall-Wice}
\affiliation{Department of Physics, Massachusetts Institute of Technology, Cambridge, MA}
\affiliation{MIT Kavli Institute, Massachusetts Institute of Technology, Cambridge, MA}

\author{Jacqueline N. Hewitt}
\affiliation{Department of Physics, Massachusetts Institute of Technology, Cambridge, MA}
\affiliation{MIT Kavli Institute, Massachusetts Institute of Technology, Cambridge, MA}

\author{Miguel F. Morales}
\affiliation{Physics Department, University of Washington, Seattle, WA}
\affiliation{Dark Universe Science Center, University of Washington, Seattle, WA} 

\author{Abraham R. Neben}
\affiliation{Department of Physics, Massachusetts Institute of Technology, Cambridge, MA}
\affiliation{MIT Kavli Institute, Massachusetts Institute of Technology, Cambridge, MA}

\author{Aaron R. Parsons}
\affiliation{Department of Astronomy, University of California Berkeley, Berkeley, CA}

\author{Haoxuan Zheng}
\affiliation{Department of Physics, Massachusetts Institute of Technology, Cambridge, MA}
\affiliation{MIT Kavli Institute, Massachusetts Institute of Technology, Cambridge, MA}

\date{October 3, 2014}


\begin{abstract}
In order to study the ``Cosmic Dawn" and the Epoch of Reionization with 21\,cm tomography, we need to statistically separate the cosmological signal from foregrounds known to be orders of magnitude brighter. Over the last few years, we have learned much about the role our telescopes play in creating a putatively foreground-free region called the ``EoR window." In this work, we examine how an interferometer's effects can be taken into account in a way that allows for the rigorous estimation of 21\,cm power spectra from interferometric maps while mitigating foreground contamination and thus increasing sensitivity. This requires a precise understanding of the statistical relationship between the maps we make and the underlying true sky. While some of these calculations would be computationally infeasible if performed exactly, we explore several well-controlled approximations that make mapmaking and the calculation of map statistics much faster, especially for compact and highly redundant interferometers designed specifically for 21\,cm cosmology. We demonstrate the utility of these methods and the parametrized trade-offs between accuracy and speed using one such telescope, the upcoming Hydrogen Epoch of Reionization Array, as a case study.
\end{abstract}

\maketitle

\section{Introduction} \label{sec:Intro}

The prospect of directly probing the intergalactic medium (IGM) during the cosmic dark ages, through the ``Cosmic Dawn" and culminating with the Epoch of Reionization (EoR) has generated tremendous excitement in 21\,cm cosmology over the past few years. Not only could it provide the first direct constraints on the astrophysics of the first stars and galaxies, but it could make an enormous new cosmological volume accessible to tomographic mapping---enabling exquisitely precise new tests of $\Lambda$CDM \cite{Yi}. For recent reviews, see e.g. \cite{FurlanettoReview, miguelreview, PritchardLoebReview, aviBook}. 

More recently, that excitement has translated into marked progress toward a statistical detection of the 21\,cm signal in the power spectrum. The first generation of experiments, including the Low Frequency Array (LOFAR \cite{LOFARinstrument}), the Donald C. Backer Precision Array for Probing the Epoch of Reionization (PAPER \cite{PAPER}), the Giant Metrewave Radio Telescope (GMRT \cite{GMRT}), and the Murchison Widefield Array (MWA \cite{TingaySummary,BowmanMWAScience}) have already begun their observing campaigns. Both PAPER \cite{DannyMultiRedshift} and the MWA \cite{X13} have released upper limits on the 21\,cm power spectrum across multiple redshifts. PAPER has already begun to use their results to constrain some models of the thermal history of the IGM \cite{PAPER32Limits}.

Still, the observational and analytical challenges that lie ahead for the field are considerable. The sensitivity requirements for a detection of the 21\,cm power spectrum necessitate large collecting areas and thousands of hours of observation across multiple redshifts \citep{MiguelNoise,Judd06,LidzRiseFall,LOFAR2,AaronSensitivity}. Of no less concern is the fact that the cosmological signal is expected to be dwarfed by foreground contaminants---synchrotron radiation from our Galaxy and other radio galaxies---by four or more orders of magnitude in brightness temperature at the frequencies of interest \citep{Angelica,LOFAR,BernardiForegrounds,PoberWedge,InitialLOFAR1,InitialLOFAR2}. 
 
The problem of power spectrum estimation in the presence of foregrounds has been the focus on considerable theoretical effort over the past few years \cite{paper1, paper2, JelicRealistic,CathWedge,LT11,DillonFast}. \citet{LT11} adapted inverse-covariance-weighted quadratic estimator techniques developed for Cosmic Microwave Background \cite{Maxpowerspeclossless} and galaxy survey \cite{Maxgalaxysurvey1} power spectrum analysis to 21\,cm cosmology. \citet{DillonFast} showed how those methods, which nominally take $\BigO{N^3}$ steps, where $N$ is the number of voxels in a 3D map or ``data cube", could be accelerated to as fast as $\BigO{N\log N}$. 
 
However, both of those works took as their starting point data cubes containing signal, foregrounds, and noise. Neither considered the important impact that an interferometer has, not just on the noise in our maps, but on the maps themselves. An instrument-convolved map or ``dirty map" has fundamentally different statistical properties than the underlying sky and the effects of the instrument cannot in general be fully undone. \citet{X13} discussed this problem approximately by assuming that point spread functions (PSFs) or ``synthesized beams" depended only on frequency. Generally speaking, that is not true; PSFs are direction-dependent and typically not invertible. In this work, we relax the assumption that went into \citet{LT11} and \citet{DillonFast} while retaining the goals they strove for: minimal information loss, rigorously understood statistics, and well-controlled approximations that make the analysis computationally feasible.
  
For any near-future 21\,cm measurement, interferometric maps are essentially an intermediate data compression step. The ultimate goal is to turn time-ordered data coming from the instrument---namely, visibilities---into statistical measurements that constrain our models of astrophysics and cosmology. So why even bother making a map if we are only going to take Fourier transforms of it and look at power spectra?  The answer to that question depends on which strategy we pursue for separating the cosmological signal from foregrounds. There are two major approaches, which we will review presently.

\begin{figure}[] 
	\centering 
	\includegraphics[width=.45\textwidth]{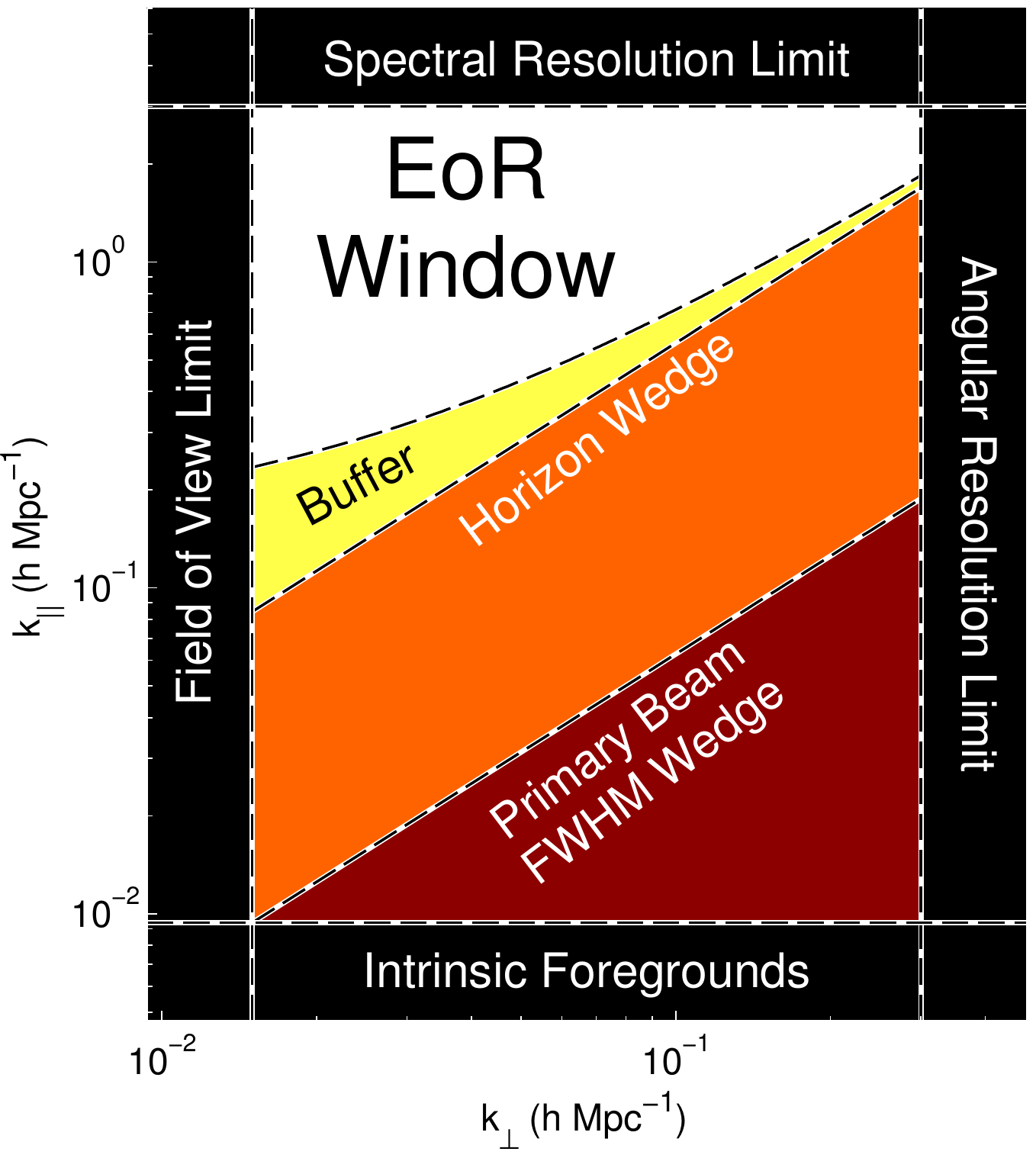}
	\caption{The ``EoR window" is a region of Fourier space believed to be essentially foreground free and thus represents a major opportunity for detecting the 21\,cm signal. Along the horizontal axis, the window is limited by the field of view, which sets the largest accessible modes, and the angular resolution of the instrument, which sets the smallest.  Along the vertical axis, the window is limited by the spectral resolution of the instrument and by the intrinsic spectral structure of galactic and extragalactic foregrounds, which dominate the spectrally smooth modes. The EoR window is further limited by ``the wedge," which results from the modulation of spectrally smooth foregrounds by the instrument's frequency-dependent and spatially varying point spread function. Much of the power in the wedge should fall below the wedge line associated with the primary beam while the horizon line serves as a hard cutoff for flat-spectrum foregrounds \cite{AaronDelay}. Limited ``suprahorizon" emission has been observed and can be attributed to intrinsic spectral structure of the foregrounds \cite{PoberWedge}, so it is possible we need a small buffer beyond the horizon to be certain that the window is foreground free. Without foreground subtraction, foregrounds are expected to dominate over the cosmological signal throughout the wedge.}
	\label{fig:EoRWindow}
\end{figure}

Over the last few years, it has been realized that a region of cylindrical Fourier space\footnote{Points in cylindrical or ``2D" Fourier space are denoted by $k_\|$, modes along the light of sight, and $k_\perp$, modes perpendicular to the line of sight. Cylindrical Fourier space takes advantage of isotropy perpendicular to the line of sight while keeping modes along the line of sight separate, since they are measured in a fundamentally different way.} should be essentially free of foreground contamination \cite{Dattapowerspec,AaronDelay,VedanthamWedge,MoralesPSShapes,Hazelton2013,CathWedge,ThyagarajanWedge,EoRWindow1,EoRWindow2}. We call this region the ``EoR window" (see Figure \ref{fig:EoRWindow}). Observations of the EoR window thus far have found it noise dominated \cite{PoberWedge,X13}. For slowly varying spectral modes (i.e. low $k_\|$), the edge of the window is set by a combination of the intrinsic spectral structure of foreground residuals and the spectral structure introduced by the instrument. Fundamentally, an interferometer is a chromatic instrument and the fact that the shape of its point spread functions depends on frequency creates complex spectral structure in 3D maps of intrinsically smooth foregrounds \cite{EoRWindow1,EoRWindow2}. 

Fortunately, there is a theoretical limit to the region of Fourier space where instrumentally induced spectral structure can contaminate the power spectrum. It is set by the delay associated with a source at the horizon (which is the maximum possible delay) for any given baseline \cite{AaronDelay}. This region of cylindrical Fourier space is known colloquially as ``the wedge." Furthermore, we expect that most of the foreground emission should appear in the main lobe of the primary beam, setting a soft limit on foreground emission at lower $k_\|$ (see Figure \ref{fig:EoRWindow}). 

The simplest approach to power spectrum estimation in the presence of foregrounds, and likely the most robust, is to simply excise the entire section of Fourier space that could potentially be foreground-dominated. This conservative approach takes the perspective that we have no knowledge about the detailed spatial or spectral structure of the foregrounds and therefore that the entire region under the wedge is hopelessly contaminated.  If that were the case, the optimal strategy would simply be to project out those modes. This ``foreground avoidance" strategy has been used to good effect by both PAPER \cite{PAPER32Limits,DannyMultiRedshift} and the MWA \cite{X13}, though neither made sensitive enough measurements to be sure that foregrounds are sufficiently suppressed inside the EoR window to make a detection without subtracting them. Considerable work has already been done with methods of estimating the power spectrum that minimize foreground contamination from the wedge into the window \cite{X13,EoRWindow2}. 

Foreground avoidance, however, comes at a significant cost to sensitivity. The more aggressive alternative is ``foreground subtraction", a strategy that tries to remove power associated with foregrounds and expand the EoR window. The idea behind foreground subtraction is twofold. First, we remove our best guess as to which part of the data is due to foreground contamination. Second, we treat residual foregrounds as a form of correlated ``noise," downweighting appropriately in the power spectrum estimator and taking into account biases introduced. In the limiting case where we know very little about the foregrounds, foreground subtraction becomes foreground avoidance. 

For the upcoming Hydrogen Epoch of Reionization Array (HERA), \citet{PoberNextGen} compared the effects of foreground avoidance to foreground subtraction. If the window can be expanded from delay modes associated with the horizon to delay modes associated with the full width at half maximum of the primary beam, the sensitivity to the EoR signal improves dramatically. Over one observing season with a 547-element HERA, the detection significance of a fiducial EoR signal improves from 38$\sigma$ to 122$\sigma$. For smaller telescopes, this might mean the difference between an upper limit and a solid detection. More importantly, the errors on the measurements of parameters that describe reionization from the power spectrum improve from about 5\% to less than 1\% when employing extensive foreground subtraction. That would be the most sensitive measurement ever made of the direct effect of the first stars and galaxies on the IGM. Simply put, there is much that might be gained by an aggressive foreground subtraction approach.

That said, it will not be easy. In order to expand the EoR window and reduce the effect of foregrounds, one must model them very carefully. Likely we will want to use outside information like high-resolution surveys to try to measure source fluxes to be much better than a percent. Even more importantly, one must take our own uncertainty about these models into account. If we do not, we risk mistakenly claiming a detection. We must propagate both our best estimates for the foregrounds and our uncertainty in our models through the instrument, which is the source of the wedge itself. 

Both galactic and extragalactic foregrounds have complex spatial structure. Any precise model for their emission is direction dependent. More importantly, our model for the statistics of our uncertainty about their emission, is also direction dependent. The covariance of residual foregrounds, especially of bright sources, is most simply and compactly expressed in real space \cite{DillonFast}.  

\begin{figure}[] 
	\centering 
	\includegraphics[width=.25\textwidth]{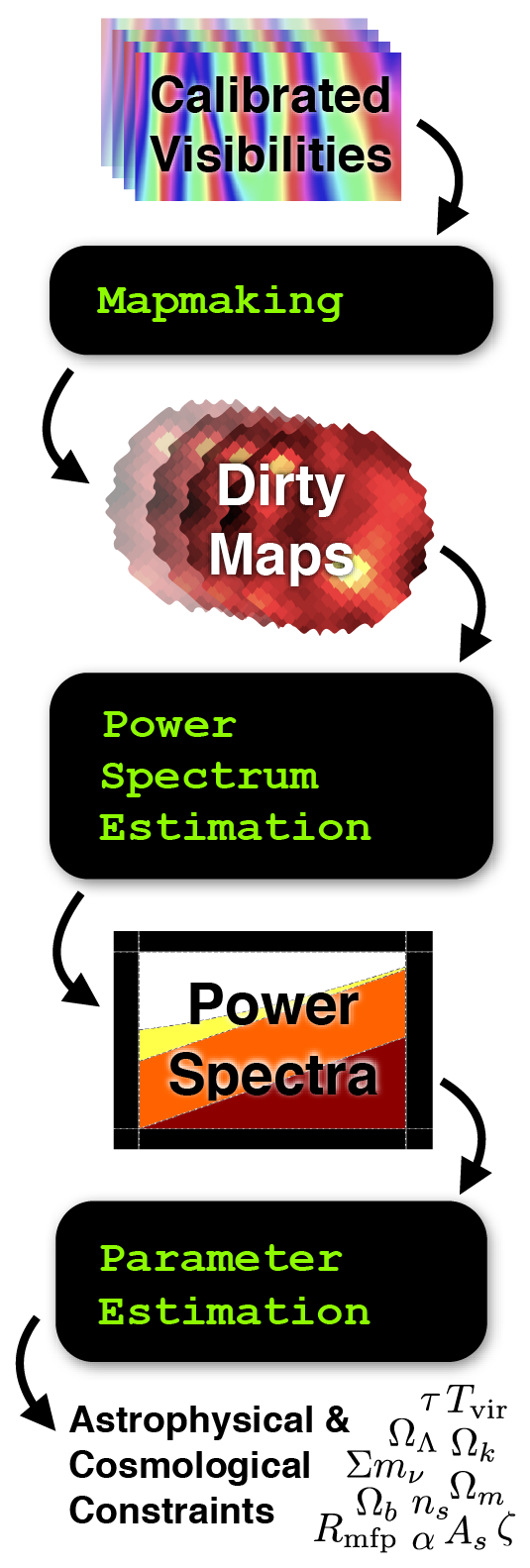}
	\caption{Mapmaking is the first in a series of steps that reduce the volume of data while trying not to lose any astrophysical or cosmological information. The goal of this work is to address that first data-compressional step---turning calibrated visibilities into a stack of dirty maps or a data cube---with any eye toward the next step---power spectum estimation in the presence of dominant astrophysical foregrounds.  This data compression is achieved by combining together different observations a single, relatively small set of maps. Power spectra represent the cosmological signal even more compactly by taking advantage of homogeneity and isotropy and serve as the natural data product to connect to simulations and theory and thus constrain cosmological and astrophysical parameters.}
	\label{fig:pipeline}
\end{figure}

We can now finally answer the question of why we should make maps if we are ultimately interested in power spectra. We need maps as an intermediate data product because they allow us to prepare our data in a highly compressed form that puts us in a natural position to carefully pick apart the signal from the foregrounds and the noise. Forming power spectra directly with visibilities, by comparison, requires treating each local sidereal time separately and vastly increases the data volume. In Figure \ref{fig:pipeline} we put mapmaking into the larger context of data reduction all the way from calibrated visibilities to cosmological and astrophysical constraints. The goal of each step is to reduce the volume of data while keeping as much cosmological information as possible, allowing for quantification of errors, and making the next step easier.

The science requirements of our maps are very different from those that motivated most interferometric mapmaking in radio astronomy to date. Usually, radio astronomers are interested in the astrophysics of what we call ``foregrounds" and focus on detailed images and spectra. For us it is especially important to understand how our maps are related statistically to the true sky, whose underlying statistics we would like to characterize using the power spectrum. Because interferometers do not uniformly or completely sample the Fourier plane, the relationship between our maps and the true sky is complicated. The PSFs of our maps depend both on frequency and on position on the sky. In order to estimate power spectra from maps accurately, we need to know precisely both the relationship of our dirty maps to the true sky and the covariance of our dirty maps that relates every pixel at every frequency to every other.\footnote{It is worth mentioning that the techniques developed here do not apply only to 21\,cm tomography. Any power spectrum made with maps produced from interferometric data needs to take into account the effects of the frequency-dependent and spatially varying PSF on both the signal and the contaminants. This includes intensity mapping of CO and CII and interferometric measurements of the CMB. Higher-order statistics, like the bispectrum and trispectrum, also need precise knowledge of the relationship between the true sky and the dirty maps.} Current imaging techniques do not compute these quantities. It is the main point of this paper to show why and how that must be done. 

Both \cite{EoRWindow1} and \cite{EoRWindow2} focused on a similar point about the important effect of the instrument on the power spectrum. There, the authors derived a framework for rigorously quantifying the errors and error correlations associated with instrument-convolved data and showed how the wedge feature arose even in a rigorous and optimal framework. However, because they formed power spectra directly from visibilities without using maps as an intermediate data-compression step, their tools are impractical for use with large data sets.

In this work, we have two main goals. First, we would like to mathematically understand how the instrument gives rise to a complicated PSF and how that PSF can be self-consistently incorporated into the inverse-covariance-weighted power spectrum estimation techniques (e.g. \cite{LT11} and \cite{DillonFast}). In Section \ref{sec:Mapmaking}, we discuss the theory of mapmaking as an intermediate step between observation and power spectrum estimation. Then, in Section \ref{sec:method}, we investigate how to put that theory into practice. We use HERA as a case study in carrying out the calculation of dirty maps and their statistics. Although the computational cost of performing those calculations is naively quite large, we develop and analyze three main ways reducing it dramatically:
\begin{itemize}
\item We explore how restricting our maps to independent facets on the sky lets us reduce the number of elements in our PSF matrices and the difficulty of calculating them (Section \ref{sec:faceting}).
\item We show how individual timesteps can be combined and analyzed simultaneously, approximately accounting for the rotation of the sky over the instrument (Section \ref{sec:snapshots}). 
\item We show how the point spread functions, while not translationally invariant,  vary smoothly enough spatially that the associated matrix operations can take advantage of certain symmetries for a computational speedup (Section \ref{sec:PSFfitting}).
\end{itemize}
We will show how each of these approximations works and analyze them to understand the trade-off between speed and accuracy in each case.

\section{Precision Mapmaking And Map Statistics in Theory} \label{sec:Mapmaking}

Making maps from interferometric data has a long history and a great number of techniques have been developed with different science goals in mind \cite{ThompsonMoranSwenson}. Most focus on deconvolution, the removal of point source side lobes (or the side lobes of extended sources represented as multiple components) after their convolution with the synthesized beam. This is the basic idea behind the CLEAN algorithm \cite{CLEAN} and its many descendants, including \cite{Schwab1984,Cotton2004,Cotton2005,Bhatnagar2008,Carozzi2009,Mitchell2008,bernardi,Bart,Smirnov2011,Li2011,FHD,WSCLEAN}. Some of these, notably that of \citet{FHD}, take inspiration from \cite{TegmarkCMBmapsWOLosingInfo}, in that they use the framework of ``optimal mapmaking" for forming dirty maps without losing any cosmological information contained in the visibilities. Additionally \cite{Richard} and \cite{ShawCoaxing}, which use the optimal mapmaking formalism in the $m$-mode basis to exploit the observational symmetries of a drift scanning interferometer, are also closely related to the work presented here.

A notable exception is \cite{SutterBayesianImaging}, which develops a method of Bayesian deconvolution via Gibbs sampling in the relatively simplified case of a gridded $uv$-plane, which can then be used for power spectrum estimation \cite{GibbsPSE}. This method  not only calculates a map but also gives error estimates on each pixel in that map. This is an especially promising technique for finding sources and quantifying the errors on our measurements of their fluxes and spectral indices. We take a different tack and do not focus on deconvolution at all.

In this work, we are interested not just in a dirty map but also in the statistical properties of that map. As in previous work, we want to know how sources are convolved with the instrument. But we also want to know how that instrumental convolution affects our covariance models for everything in the map, including signal, noise, and foregrounds. A complete understanding of the relationship between the true sky and our dirty maps will allow us to comprehensively model these important statistical quantities. Current imaging methods simply do not compute that relationship and the resulting noise covariance matrix. However, these are required for methods of power spectrum estimation in order to properly weight data in the presence of correlated noise and foregrounds and to account for missing modes. The importance of this was realized by \cite{FFTT2}, though we will use a different computational approach to speed up the calculations. 

We begin this section by summarizing the relevant physics behind interferometry in Section \ref{sec:interferometry}. We then review the optimal mapmaking formalism in Section \ref{sec:OMM}. Finally, in Section \ref{sec:powerspectra} we work out the consequences of proper map statistics for the inverse-covariance-weighted quadratic power spectrum estimation formalism, including how they affect the models of the covariance of cosmological signal, noise, and foreground residuals.

\subsection{Interferometric Measurements} \label{sec:interferometry}

When we make maps from interferometric data, we are interested in computing a map estimator or ``dirty map," which we call $\widehat{\x}$, and understanding its relationship to $\x$, the true, discretized sky.\footnote{We write these quantities as vectors as a compact way of combining indices over both angular dimensions on the sky and over frequency.} We do not have access to $\x$ directly; we can only make inferences about it by making a set of complex ``visibility" measurements which we call $\y$. Each measurement made with our instrument is a linear combination of the true sky added to instrumental noise. Therefore, we can represent all our measurements with
\beq
\y = \A \x + \n, \label{eq:measurement}
\eeq
where $\A$ represents the interferometric response of our instrument over all times, frequencies, and baselines and where each $n_i$ is the instrumental noise on the $i$th visibility. The matrix $\A$ has the dimensions of the number of measured visibilities (for every baseline, frequency, and integration) by the number of voxels in the 3D sky (all pixels at all frequencies).

The statistics of $\n$ are fairly simple. It has zero mean and the noise on each visibility is generally treated as independent of that on every other visibility. Therefore,
\begin{align}
\langle n_i \rangle &= 0 \\
N_{ij} \equiv \langle n_i n_j^* \rangle &= \sigma^2_i \delta_{ij}.
\end{align}
The form of $\A$ is considerably more complicated, it can be written in the form of Equation \eqref{eq:measurement} because a visibility is a weighted integral over the whole sky which can be approximated to any desired precision by a finite matrix operation.

The visibility measured by a noise-free instrument with arbitrarily fine frequency resolution at frequency $\nu$ and baseline $\base_m$ in response to a sky specific intensity $I(\rhat,\nu)$ defined continuously over all points on the sky $\rhat$ is
\begin{align}
V(\base_m,\nu) = \int B_m\left(\rhat,\nu \right)
I(\rhat,\nu) \exp\left[-2 \pi i  \frac{\nu}{c} \base_m \cdot \rhat \right] d\Omega . \label{eq:AnalyticVisibility}
\end{align}
Here $B_m(\rhat,\nu)$ is the product of the complex primary beams of the two antenna elements that form the $m$th baseline. In this equation and in the rest of this section, we will ignore the polarization of the sky and the fact that there are different beams for each polarization, assuming homogenous antenna elements. We do this for simplicity; the results are straightforwardly generalizable to a complete treatment of polarization, which we will explore in  Appendix \ref{app:PolAndBeams}. In that appendix, we will also look at how heterogenous arrays straightforwardly incorporated into our framework as well.

Given a finite number of measurements, we are interested in the relationship between visibilities and a discretized true sky, $\x$. In frequency, that discretization comes from the spectral response of our instrument---we can only measure a limited number of frequency channels. Spatially, we need to choose our pixelization of the sky. Let us define a 3D pixelization function $\psi_i(\rhat,\nu)$ that incorporates both these kinds of pixelization. It is defined so that,
\beq
x_i = \int  \psi_i(\rhat,\nu) \frac{c^2}{2k_B\nu^2} I(\rhat,\nu) d\Omega d\nu ,\label{eq:PixelizationDefinition}
\eeq
where the extra factor of $c^2/2k_B\nu^2$ converts from units of specific intensity to brightness temperature. For simplicity, we define $\psi_i(\rhat,\nu)$ to be the unitless top-hat function, normalized such that  
\beq
\int \psi_i(\rhat,\nu) \frac{d\Omega}{\Delta \Omega} \frac{d\nu}{\Delta \nu} = 1 \label{eq:pixelization_defintiion}
\eeq
where $\Delta \nu$ is the frequency resolution of the instrument and $\Delta \Omega$ is the angular size of the pixels. Other choices of $\psi_i(\rhat,\nu)$ are perfectly acceptable, in which case $\Delta \nu$ and $\Delta \Omega$ become characteristic spectral and spatial sizes of pixels.

Therefore we can rewrite Equation \eqref{eq:AnalyticVisibility} as a sum:
\begin{align}
V(\base_m,\nu_n) \approx \sum_k &\Delta\Omega \frac{2 k_B \nu_n^2}{c^2} x_k(\nu_n) \times \nonumber \\
&B_m(\rhat_k,\nu_n) \exp\left[-2 \pi i  \frac{\nu_n}{c} \base_m \cdot \rhat_k \right]. \label{eq:VisibilitySum}
\end{align}
Here we have chosen to break apart the index $i$ into a spatial subindex, $k$, and a spectral subindex, $n$. The sum is over all spatial pixels. This approximation relies on choosing a frequency and angular resolution small enough that $B(\rhat,\nu)$ and $ \exp\left[-2 \pi i  (\nu / c) \base_m \cdot \rhat \right]$ can be approximated as constants inside of a single spatial pixel and frequency channel. Since $V(\base_m,\nu_n)$ is an entry in $\y$, Equation \eqref{eq:VisibilitySum} gives us the elements of $\A$ by relating $\y$ to $\x$ for a single observation and a single baseline. Of course, the full matrix $\A$ that goes into Equation \eqref{eq:measurement} gives us a relationship between the true sky and every visibility at every frequency and at every local sidereal time. The basic physics, however, is captured by Equation \eqref{eq:VisibilitySum}.

\subsection{The Optimal Mapmaking Formalism}\label{sec:OMM}

Given a set of visibilities (or any time-ordered data) of the form in Equation \eqref{eq:measurement}, there is a well known technique for forming estimators of the true sky without losing any information about the discretized sky contained in the time-ordered data \cite{TegmarkCMBmapsWOLosingInfo}. Those estimators, known as ``optimal mapmaking" estimators, take the general form
\beq
\widehat{\x} = \D \A^\dagger \N^{-1} \y \label{eq:OMM}
\eeq
where $\D$ can be any invertible normalization matrix. Especially for long observations, $\y$ is a much larger vector than  $\widehat{\x}$. Mapmaking represents a major data compression step. 

\begin{figure*}[] 
	\centering 
	\includegraphics[width=1\textwidth]{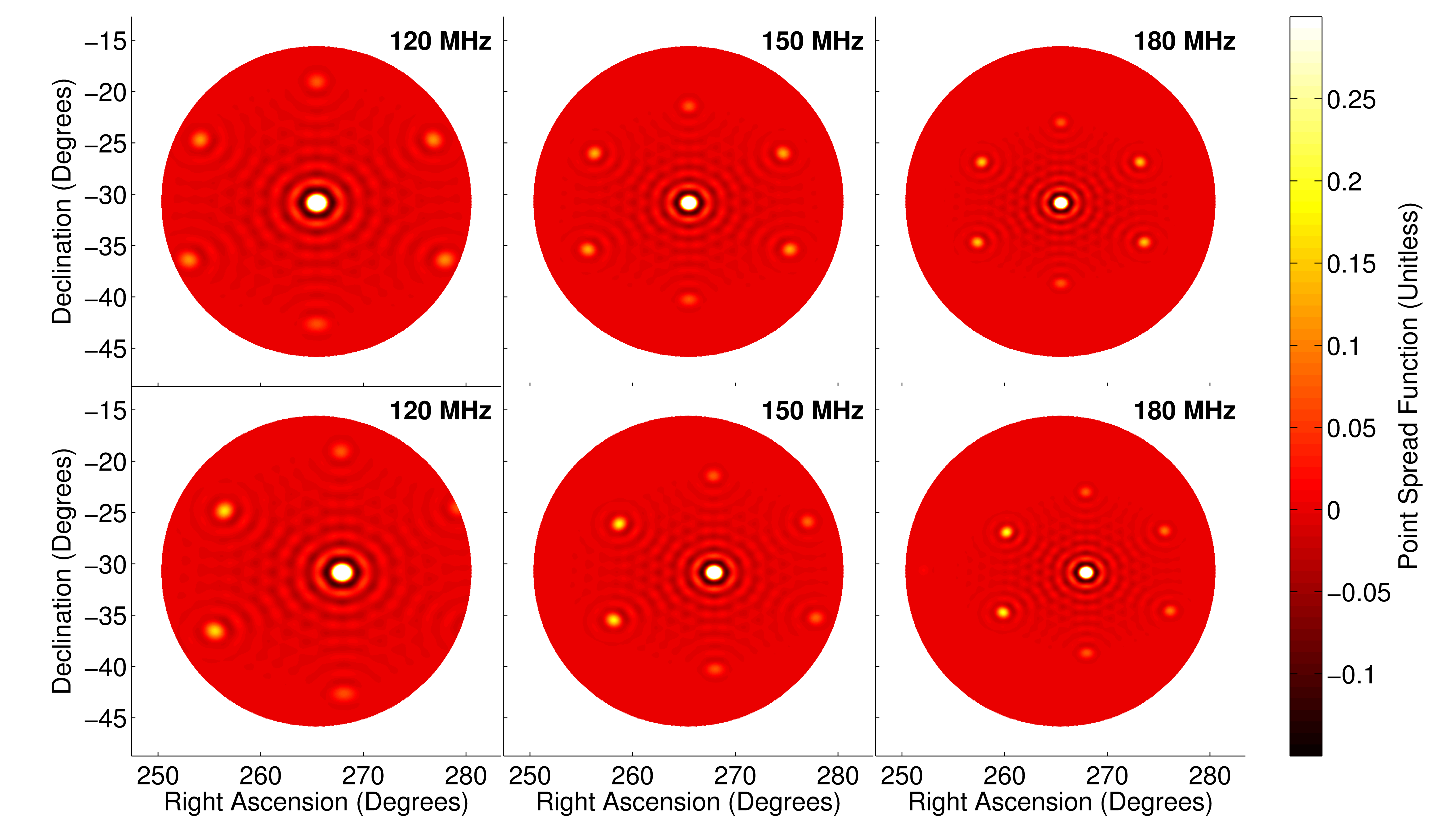}
	\caption{The point spread function (or equivalently, the synthesized beam) of a dirty map varies both as a function of position on the sky and as a function of frequency. In the top row, we show the point spread functions at three frequencies corresponding to the center of the primary beam calculated for HERA. They exhibit clear diffraction rings and fairly strong side lobes due to tje fact that the minimum separation between antennas is significantly longer than the wavelength. The hexagonal pattern is due to the geometry of the array. In the bottom row, we look at off-center point spread functions. These also have side lobes, though they are asymmetric due to the primary beam and the projected layout of the array and thus a clear example of the translational variation of the PSF. All six can be thought of as single rows of different frequency blocks of the full matrix of point spread functions, $\PSF$. Each PSF peaks at 1, but we have saturated the color scale to show detail. In Section \ref{sec:method}, we will explain in detail how these PSFs are calculated.}
	\label{fig:PSFs}
\end{figure*}

The expected value of the estimator is
\begin{align}
\langle \widehat{\x} \rangle &= \langle \D \A^\dagger \N^{-1} (\A \x + \n) \rangle \nonumber \\ 
&= \D \A^\dagger \N^{-1} (\A \x + \langle \n \rangle) \nonumber \\ 
&= \D \A^\dagger \N^{-1} \A \x. \label{eq:<xhat>}
\end{align}
In general, the expected value of $\widehat{x}$ is not the same as the true sky but is rather some complicated linear combination of pixels on the true sky. We define 
\beq
\PSF \equiv \D \A^\dagger \N^{-1} \A \label{eq:PSFdef}
\eeq
to be the matrix of point spread functions. Each column of this matrix tells us how each pixel on the true sky gets mapped to all the pixels of the dirty map. If we want to normalize the PSF to always have a central value of 1, we can achieve that by a judicious choice of $\D$. In this work, we make that choice of PSF normalization. Recall that $\D$ can be any invertible matrix. Since we are not trying to make images that look as much as possible like the true sky but rather just to keep track of exactly how our dirty maps are related to the true sky, making a very simple choice for $\D$ is sensible.\footnote{The choice of $\D=\left[\A^\dagger \N^{-1} \A\right]^{-1}$  was used by WMAP \cite{WMAPconjugategrad} because it makes  $\PSF = \Eye$, but that matrix is generally not invertible in radio interferometry. Whenever one cannot make that choice of $\D$, $\PSF$ is not the identity and one must keep track of its effects.} Therefore, we use our freedom in choosing $\D$ to make it a diagonal matrix---effectively a per-pixel normalization. In Figure \ref{fig:PSFs} we plot an example of the central portions of two different rows of $\PSF$ at three different frequencies. 

\subsection{Connecting Maps to Power Spectra}\label{sec:powerspectra}

As we discussed earlier, we are interested in mapmaking in order to reduce the volume of our data without losing any sky information or the ability to remove foregrounds. From the map, the next step is to further compress the data by calculating a power spectrum, which can be directly compared with theoretical predictions. To connect the mapmaking formalism to 21\,cm power spectrum estimation, we will review the statistical estimator formalism for calculating power spectra while not losing any cosmological information. In the process, we will enumerate the quantities that we need to calculate in order to estimate a power spectrum from $\widehat{\x}$. Then we will show the form that those quantities take in terms of $\widehat{\x}$, $\PSF$, and $\D$.

\subsubsection{Power Spectrum Estimation Reivew}

Fundamentally, a power spectrum estimate is a quadratic combination of the data. To calculate a power spectrum, roughly speaking, one simply Fourier transforms real-space data, squares, and then averages in discrete bins to form ``band powers." In a real-world measurement with noise and foreground contamination, we need a more sophisticated technique. 

Because we have a finite amount of data, we must discretize the power spectrum we estimate by approximating $P(\mathbf{k})$ as a piecewise constant function described by a set of band powers $\mathbf{p}$ using 
\beq
P(\mathbf{k}) \approx \sum_\alpha p_\alpha \chi_\alpha(\mathbf{k}), \label{eq:bandpowers}.
\eeq
Here $\chi_\alpha(\mathbf{k})$ is a characteristic function which equals 1 inside the region described by the band power $p_\alpha$ and vanishes elsewhere.

Since the power spectrum is a quadratic quantity in the data, an estimator $\widehat{\p}$ of the band power spectrum $\p$ (which is discretized by approximating the power spectrum as piecewise-constant) takes the form
\beq
\widehat{p}_\alpha = (\widehat{\x} - \mean)^\trans \mathbf{E}_\alpha (\widehat{\x} - \mean) - b_\alpha.
\eeq
Here $\E_\alpha$ very generally represents the operations we want to perform on the data and $\mean \equiv \langle \widehat{\x} \rangle$ is the ensemble average over many realizations of the same exact observation, each with different noise, and $\mathbf{b}$ removes additive bias from noise and residual foregrounds in the power spectrum.

Just as estimators of the form in Equation \eqref{eq:OMM} do not lose any information about the true sky contained in the visibilities, there exists an optimal quadratic estimator for power spectra that does not lose cosmological information \cite{Maxpowerspeclossless}.\footnote{This entails certain assumptions, most notably that the noise, residual foregrounds, and signal are all completely described by their means and covariances---in other words that they are Gaussian. We know that this is not exactly true in the case of residual foregrounds and signal, though it is generally assumed to be a pretty good approximation for the purposes of the first generation of 21\,cm measurements \cite{LT11}.} Those estimators take the form
\beq
\widehat{p}_\alpha = \frac{1}{2}M^{\alpha\beta} (\widehat{\x} - \mean)^\trans \C^{-1} \mathbf{C},_\beta \C^{-1} (\widehat{\x} -\mean) - b_\alpha. \label{eq:QE}
\eeq
In this equation, $\M$ is an invertible normalization matrix, analogous to $\D$ and $\C$ is the covariance of $\widehat{\x}$ (not of the true sky $\x$) and is defined as 
\beq
\C \equiv \langle \widehat{\x} \widehat{\x}^\trans \rangle - \langle \widehat{\x} \rangle \langle \widehat{\x} \rangle ^\trans.
\eeq
Each $\mathbf{C},_\beta$ matrix, which encodes the Fourier transforming and binning steps of the power spectrum, is defined such that
\beq
\C = \C^\text{contaminants} + \sum_\beta p_\beta \mathbf{C},_\beta.
\eeq
Here $\C^\text{contaminants}$ represents the covariance of anything that appears in $\widehat{\x}$ that is not the 21\,cm cosmological signal. In other words, the set of $\mathbf{C},_\beta$ matrices tells us how the covariance of $\widehat{\x}$ responds to changes in the underlying band powers, $\p$. We will explain the precise form of $\C,_\beta$ shortly.

\subsubsection{The Statistics of the Mapmaking Estimator} 

All of the quantities we are interested in calculating when estimating the power spectrum, including the bias term, the errors on our band powers, the error covariance between band powers, and the ``window functions" that encode the relationship between $\widehat{\p}$ and $\p$, are derived from our models of $\mean$ and $\C$ (see e.g. \cite{Maxpowerspeclossless, LT11, DillonFast, X13} for the exact forms of these quantities). In this section, we will see how those quantities depend on the mapmaking algorithm and are inextricably linked to the response of the interferometer. 

We have already shown that $\langle \widehat{\x} \rangle = \PSF \x$ in Equations \eqref{eq:<xhat>} and \eqref{eq:PSFdef}. When we are making a map, this is sufficient---there is a ``true" sky and we are trying to estimate a quantity related to it from noisy data in a well-understood way. In the context of power spectrum estimation, simply averaging down instrumental noise is not enough. Because we are interested in the statistical properties of the Universe as a whole, we are trying to use multiple independent spatial modes to learn about at the underlying statistics of $\x$, taking advantage of homogeneity and isotropy. Though there is only one true sky, we treat it as a random field with Gaussian statistics. Therefore,
\begin{align}
\mean &= \langle \widehat{\x} \rangle  = \PSF \langle \x \rangle \nonumber \\ 
&= \PSF\left[ \langle \x^S \rangle + \langle \x^N \rangle + \langle \x^{FG} \rangle  \right] = \PSF  \langle \x^{FG} \rangle. \label{eq:mean}
\end{align}
Here we have explicitly separated our model for the sky into three statistically independent parts: the 21\,cm signal, the noise, and the foregrounds. Only the foregrounds have nonzero mean.\footnote{The mean of the cosmological signal is zero only because it is usually defined as the fluctuations from the mean brightness temperature of the global 21\,cm signal. For our purposes, the global signal is a contaminant and can be treated as part of the diffuse foregrounds without loss of generality.} Because they are statistically independent, the covariance can be separated into the sum of three matrices.\footnote{It should be noted that each of these covariance matrices is the covariance of the instrument-convolved sky and not the true sky, in contrast to the notation in \cite{DillonFast} which, by treating an idealized scenario, ignored the distinction.} Hence,
\begin{align}
\C = \C^S + \C^N + \C^{FG}.
\end{align}
We will now show how all of these are calculated in the context of optimal mapmaking.

\subsubsection{The Signal Covariance}

First, let us turn to the signal covariance, $\C^S$. To understand what this really means, we need to first explain what we mean by $\x^S$. Imagine a continuous 21\,cm temperature field as a function of position in comoving coordinates, $x^S(\mathbf{r})$. Each element of the vector $\x^S$ is given by
\beq
x^S_i \equiv \int \psi_i(\mathbf{r})x^S(\mathbf{r}) \frac{d^3r}{\Delta V}, 
\eeq
where $\psi_i(\mathbf{r})$ encloses exactly the same volume as $\psi_i(\rhat,\nu)$ and $\Delta V \equiv \int \psi_i(\mathbf{r}) d^3r$ is the comoving volume of a voxel. The continuous 21\,cm power spectrum, $P(\mathbf{k})$ is defined by
\beq
\left< \left[\widetilde{x}^S(\mathbf{k})\right]^* \widetilde{x}^S(\mathbf{k}') \right> \equiv (2\pi)^3 \delta(\mathbf{k} - \mathbf{k}') P(\mathbf{k}), 
\eeq
where $\widetilde{x}^S(\mathbf{k})$ is the Fourier transform of $x^S(\mathbf{r})$. It follows then that
\beq
\langle x^S_i x^S_j \rangle - \langle x^S_i \rangle \langle x^S_j \rangle = \int \widetilde{\psi}_{i}(\mathbf{k})\widetilde{\psi}_{j}^{*}(\mathbf{k})P(\mathbf{k})\frac{d^3k}{(2\pi)^{3}}. \label{eq:signalcov}
\eeq

By combining Equations \eqref{eq:signalcov} and \eqref{eq:bandpowers}, we can write down the covariance of $\mathbf{x}^S$:
\beq
\langle x^S_i x^S_j \rangle - \langle x^S_i \rangle \langle x^S_j \rangle \approx \sum_\alpha p_\alpha Q^\alpha_{ij}, \label{eq:covx}
\eeq
where
\beq
Q^\alpha_{ij} \equiv \int \widetilde{\psi}_{i}(\mathbf{k})\widetilde{\psi}_{j}^{*}(\mathbf{k})\chi_\alpha(\mathbf{k})\frac{d^3k}{(2\pi)^{3}}.
\eeq
Finally, using the fact that $\langle \widehat{\x} \rangle = \PSF \x$ determines also the relationship between the cosmological components of $\x$ and $\widehat{\x}$, we find that
\beq
\C^S \approx \PSF \left[ \sum_\alpha p_\alpha \Q_\alpha \right]\PSF^\trans \label{eq:CS}
\eeq
and therefore that 
\beq
\C,_\alpha \approx \PSF \Q_\alpha \PSF^\trans. \label{eq:CcommaAlpha}
\eeq

\subsubsection{The Noise Covariance}
While $\langle \widehat{\x}^N \rangle = \langle \x^N \rangle = 0$, the instrumental noise still contributes to the covariance. Our mapmaking formalism makes it straightforward to track how the noise on individual visibilities, $\sigma^2_i$, translates into correlated noise between pixels in a dirty map, which is described by $\C^N$. Let us imagine that $\x = 0$ and our instrument measured just noise for each visibility. If we compute the covariance of $\widehat{\x}$ in this case we will have $\C^N$, since $\C^{S}$ and $\C^{FG}$ represent our knowledge about the sky. This is true because there are no cross terms that correlate noise with foregrounds or signal.

Therefore, since our usual inverse-covariance-weighted map estimator now gives us
\beq
\widehat{\x}^N = \D \A^\dagger \N^{-1} \n,
\eeq
it follows that
\begin{align}
\C^N &= \left< \widehat{\x}^N \left(\widehat{\x}^N \right)^\trans \right> =  \left< \D \A^\dagger \N^{-1} \n \n^\dagger \N^{-1} \A \D^\trans \right> \nonumber \\
&= \D \A^\dagger \N^{-1} \left< \n \n^\dagger \right> \N^{-1} \A \D^\trans \nonumber \\ 
&= \D \A^\dagger\N^{-1} \A \D^\trans = \PSF \D^\trans. \label{eq:CN}
\end{align}
This is a gratifyingly simple result; calculating $\PSF$ yields $\C^N$ virtually for free. It also allows us to avoid the common assumption (made for example by \cite{X13}, \cite{LT11} and, \cite{DillonFast}) that instrumental noise is uncorrelated between pixels in a gridded $uv$-plane. Correlations between $uv$ pixels introduced by the primary beam are fully taken into account in our framework because, like in \cite{EoRWindow1}, $\C^N$ contains all the relevant information about the instrument and the mapmaking process.

\subsubsection{The Foreground Covariance}
Finally, we come to the statistics of the foregrounds. The reason that we treat $\x^{FG}$ as a random field even though there is really only one set of true foregrounds is that we want to represent both our best guess at the foregrounds and our uncertainty about that guess. When we write $\langle \x^{FG} \rangle$ in Equation \eqref{eq:mean}, we really mean our best guess as to the true foregrounds---the average of our incomplete knowledge about their positions, fluxes, spectral indices, and angular extents. Therefore we need to calculate
\beq
\mean = \langle \widehat{\x}^{FG} \rangle = \PSF \langle \x^{FG} \rangle = \PSF \x^{FG}_\text{model}
\eeq
to use in our quadratic estimator in Equation \eqref{eq:QE}.
 
Previous work (e.g.  \cite{LT11,DillonFast}) built explicit models of the foreground uncertainty by looking at the first and second moments of $\x^{FG}$ and not at $\widehat{\x}^{FG}$. We can take that work and generalize it straightforwardly. If $\C^{FG}_\text{model}$ is a model of foregrounds that takes into account our uncertainties about fluxes, spectral indices, and angular correlations, like the one developed in \cite{LT11} and \cite{DillonFast}, then the foreground covariance of the estimator is
\beq
\C^{FG} = \PSF \C^{FG}_\text{model} \PSF^\trans. \label{eq:CFG}
\eeq

\begin{figure}[] 
	\centering 
	\includegraphics[width=.5\textwidth]{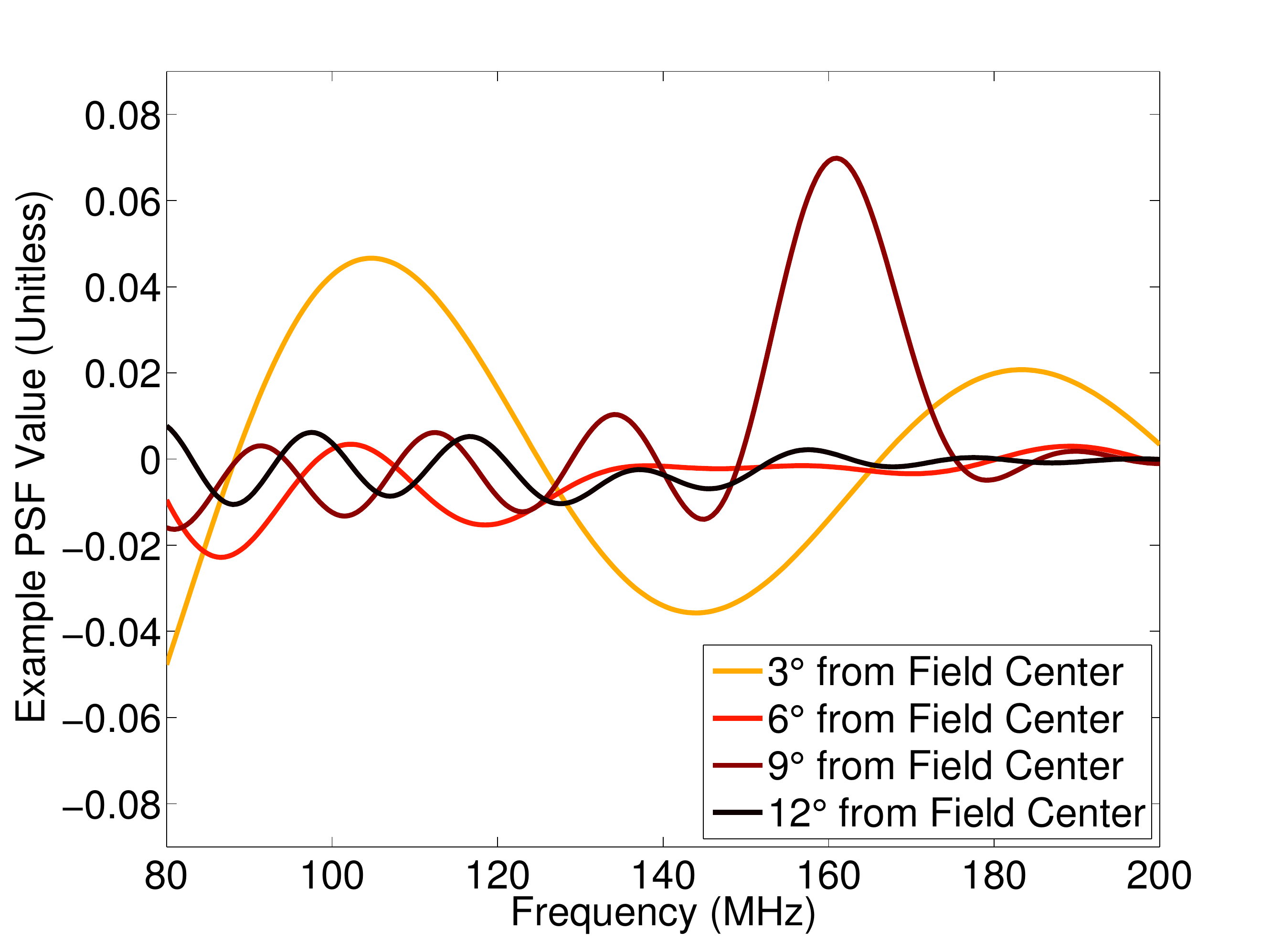}
	\caption{The position and frequency dependence of the synthesized beam is the origin of the ``wedge" feature and plays a key role in determining which Fourier modes are foreground dominated in any power spectrum estimate. Here we show four different example lines of sight through a single frequency-dependent PSF, namely the one we showed for HERA in the top row of Figure \ref{fig:PSFs}. The structure we see means that intrinsically flat spectrum sources will appear far more complicated in a dirty map. We can also see that emission further from the zenith has more complicated spectral structure---an observation that helps explain the wedge. Any attempt at foreground subtraction will require detailed knowledge of this spectral behavior, both for our models for foregrounds and for our models of our uncertainty about foreground fluxes and spectral indices.}
	\label{fig:chromaticity} 
\end{figure}

This equation compactly illustrates a key difference between the analysis methods developed by \citet{LT11} and \citet{DillonFast} and any future work that takes into account the inherent frequency dependence of foregrounds in dirty maps---the focus of this work. Intrinsic foregrounds are believed to be dominated by only a few Fourier modes \cite{AdrianForegrounds}. That means that the expression of our uncertainty about the level of foreground contamination and thus our ability to subtract foreground, $C^{FG}_\text{model}$, should also be dominated by a few Fourier modes. However the PSF's spectral and spatial structure moves power from those low $k_\|$ modes up into the wedge. In Figure \ref{fig:chromaticity}, we plot a few representative lines of sight of a field-centered PSF of a zenith-pointed instrument at different distances from field center. 
Even a flat-spectrum source would see considerable structure introduced on many spatial scales along the line of sight, especially far from the zenith. This is the origin of the wedge \cite{MoralesPSShapes} and, as \cite{EoRWindow1} pointed out, it can be fully understood as a consequence of the fact that frequency appears in the exponent of Equation \eqref{eq:AnalyticVisibility}. An interferometer is an inherently chromatic instrument. 

To summarize, in order to optimally estimate a 21\,cm power spectrum from the results of an optimal mapmaking routine, we must properly take into account the relationship between the dirty map and the true sky. To do this, we will need:
\begin{enumerate}
\item Our estimated dirty map, $\widehat{\x}$.
\item The normalization matrix for that map, $\D$, and the matrix of point spread functions, $\PSF$. Those require knowledge of the instrument, the observing strategy, and the noise in our measurements.
\item A model for the cosmological signal, which will allow us to properly account for sample variance.
\item A ``best guess" for the foregrounds and a model for our uncertainty about that best guess.
\end{enumerate}
With all these components, we can go from visibilities, through the data-compressing mapping step, and all the way to band powers in a self-consistent way while minimizing the loss of cosmological information and maintaining a full understanding of the error properties of our measurements.

\section{Precision Mapmaking in Practice: Methods, Trade-Offs, and Results} \label{sec:method}

The theoretically optimal mapmaking method outlined in Section \ref{sec:Mapmaking} poses immense computational challenges. To make it useful for real-world application, we need to find and assess ways of simplifying it while maintaining its precision and statistical rigor.

Because this work serves in large part to generalize the work of \cite{DillonFast}, it is essential to continue to assess that the proposed algorithms are computationally feasible, despite the large size of these data sets and the potentially cost-prohibitive matrix operations involved. That work showed that as long as $\C$ could be decently preconditioned and then multiplied by a vector quickly, we could estimate the power spectrum in a way that scaled favorably with the data volume---between $\BigO{N\log N}$ and $\BigO{N^{5/3}}$, where $N$ is the number of voxels in a data volume. This was accomplished using various numerical tricks, taking advantage of translational invariance, the fast Fourier transform, various symmetries, and the flat-sky approximation.

Without any approximations, the vectors and matrices we introduced in Section \ref{sec:Mapmaking} are very big. $\PSF$, for example relates the whole true sky to the whole dirty map---for every frequency, it has as many entries as the number of pixels squared. The time-ordered data vector is very big too---it has entries for every baseline, at every frequency, for every integration. That means that $\A$ is enormous, since it maps from $\x$ to $\y$. We quantify exactly the exact scale of the problem of data volume and computational difficulty in Section \ref{sec:challenges}, but it is clear that calculating every vector and matrix quantity we have enumerated in Section \ref{sec:Mapmaking} is not feasible.

When making maps, there are at least six ways to make $\widehat{\x}$ and $\PSF$ smaller or easier to calculate or use. Three have to do with the geometry of $\widehat{\x}$; three have to do with approximate methods of calculating $\widehat{\x}$ or $\PSF$:
\begin{enumerate}
\item We can make faceted maps of only very small parts of the sky at a time.
\item We can pixelize the sky more coarsely.
\item We can average together neighboring frequencies, lowering the frequency resolution.
\item We can average together neighboring timesteps before computing $\PSF$.
\item We can make $\PSF$ smaller by taking advantage of the finite sizes of the primary and the synthesized beams.
\item We can make $\PSF$ sparser by approximately fitting it in some basis.
\end{enumerate}
Roughly speaking, the first three approaches affect the kind of maps we want to make and the information content in them. The last three affect the quality of the maps we make or the fidelity with which an approximate version of $\PSF$ represents the relationship between $\widehat{\x}$ and $\x$. The exact properties of the desired maps depends upon the power spectrum estimation technique used. For example, if we want to measure high $k_\perp$ modes, we need high angular resolution and therefore a lot of pixels.

In this work, we take a specific case of the first three---choices motivated by the particular array we assess and the desire not to lose much cosmological information. We then evaluate quantitatively the trade-offs inherent in approaches that affect the quality of $\widehat{\x}$ and any approximation to $\PSF$. We begin by specifying both the array (Section \ref{sec:HERA}) and the sky model (Section \ref{sec:skymodel}) that we use for the case study we present. In that context, we can quantify the computational challenges involved in mapmaking in Section \ref{sec:challenges}. 

From there, we examine the three ways of making the mapmaking problem easier for a given kind of map. In Section \ref{sec:faceting} we look at truncating $\PSF$ and how that affects our understanding of the relationship between the dirty map and the true sky. In Section \ref{sec:snapshots} we look at the optimal way to perform time averaging and the trade-offs involved. Then we look at finding a sparse approximation to $\PSF$ in Section \ref{sec:PSFfitting}, which is important because multiplication by all three parts of $\C$ also requires multiplication by $\PSF$. We discuss a way of accomplishing that in the spirit of \cite{DillonFast}.\footnote{The question of preconditioning for rapid conjugate gradient convergence, which was addressed in \cite{DillonFast} in the context of estimators based on $\x$ rather than $\widehat{\x}$, is left for future work. That question cannot be answered until the exact form of the $\widehat{\x}$ is chosen. We may choose estimators with a tapering function, such as those suggested by \cite{EoRWindow1} and \cite{EoRWindow2}. We may also choose to project out certain modes from the dirty map, as we discuss in Appendix \ref{app:projection}.} All of these speed-ups require small approximations and we  assess the effect of those approximations quantitatively. Finally, in Section \ref{sec:computationalSummary} we summarize those results and what we can confidently say so far about the accuracy requirements for approximating $\widehat{\x}$ and $\PSF$ for the purposes of 21\,cm power spectrum estimation.
 
\subsection{HERA: A Mapmaking Case Study} \label{sec:HERA}

\begin{figure}[] 
	\centering 
	\includegraphics[width=.48\textwidth]{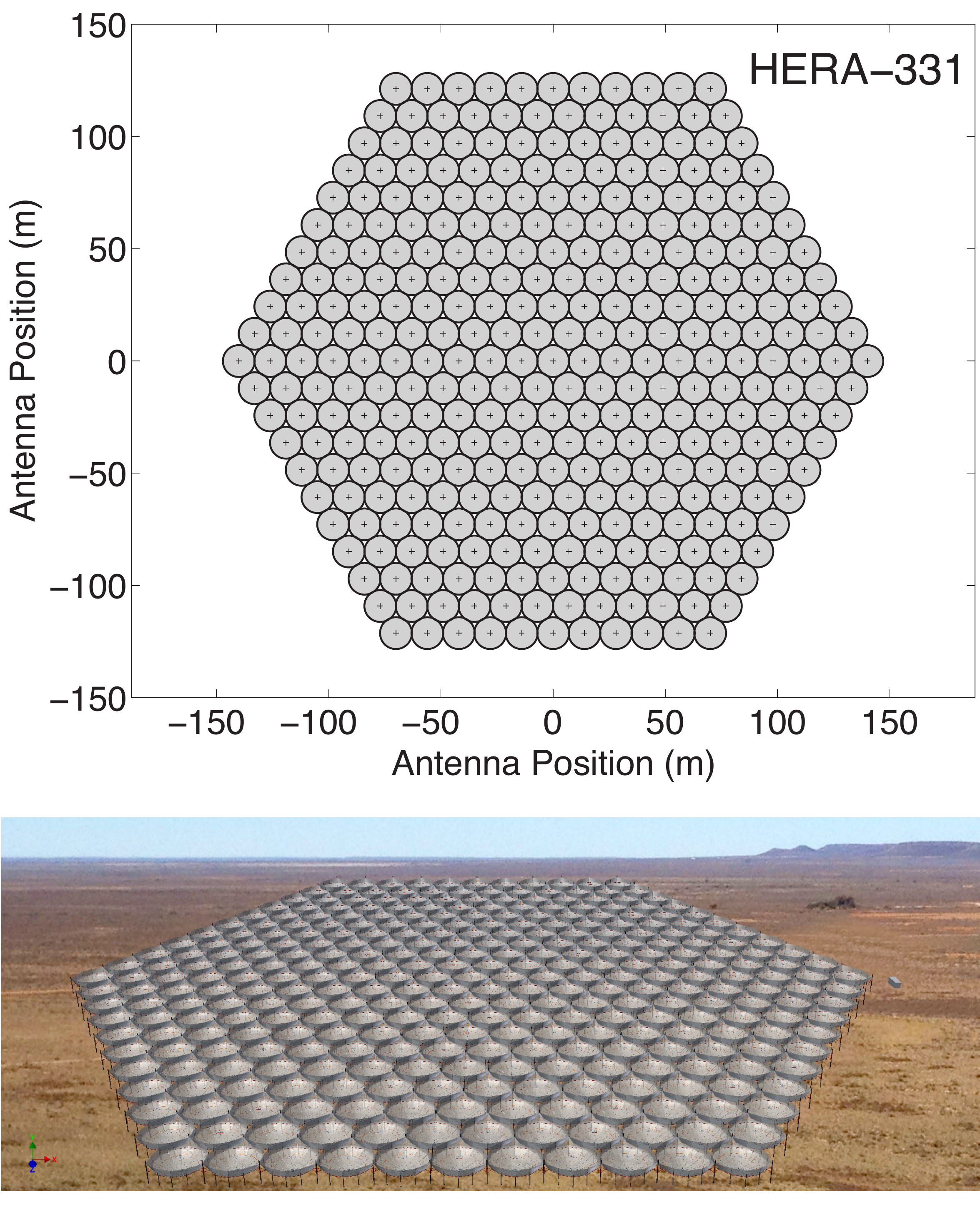}
	\caption{We test our method on simulated visibilities from the planned Hydrogen Epoch of Reionization Array (HERA). The array, seeen schematically in the top panel, consists of 331 14\,m parabolic dishes, arranged in a close-packed hexagonal configuration. In the bottom panel, we show a rendering of the final array, which will feature more than $0.05\text{\,km}^2$ of collecting area (a standard shipping container, on the right side of the image, is shown for comparison.)}
	\label{fig:HERA-331}
\end{figure}

To test our mapmaking method and our techniques for speeding it up, we need to simulate the visibilities that a real instrument would see. We choose the planned design of the recently commenced Hydrogen Epoch of Reionization Array (HERA) as a particularly timely and relevant case study. HERA will have 331 parabolic dishes, each 14\,m in diameter. They will be fixed to point at the zenith with crossed dipole antennas suspended at prime focus. They will be arranged into a maximally dense hexagonal packing (see Figure \ref{fig:HERA-331}), both to maximize sensitivity to cosmological modes \cite{AaronSensitivity,PoberNextGen} and for ease and precision of calibration \cite{redundant,MITEoR_IEEE,MITEoR}.\footnote{Plans for HERA also include outrigger antennas at much greater distances from the hexagonal core to enable low signal-to-noise, high angular resolution imaging. Though they will be useful for making high-resolution maps and modeling astrophysical foregrounds, they do not add significantly to the cosmological sensitivity of the instrument. Since we are focused on maps as a data-compression step between visibilities and power spectra, we ignore them in this analysis.} In this work, our calculations assume perfect calibration of the instrument and (unless otherwise stated) perfect antenna placement.

HERA also has two advantages that make our algorithms easier to carry out on a relatively small number of computers. First, although it has 331 elements, it only has 630 unique baselines. That is because a highly-redundant array with $N$ baselines has $\BigO{N}$ unique baselines, as opposed to minimally redundant arrays, which have $\BigO{N^2}$ baselines. That is why the MWA has an order of magnitude more baselines than HERA, even though it has only 128 elements. Second, it has a relatively small primary beam, in contrast to both MWA and PAPER. In this work, we model it fairly accurately as a Gaussian beam with a full width at half maximum of $10^\circ$ at 150\,MHz. It should be noted that the method described in this work is independent of the interferometric design. HERA happens to be both a particularly convenient and relevant example.

\subsection{Testing Mapmaking with a Specific Sky Model} \label{sec:skymodel}

As we find ways to compute mapmaking statistics quickly and accurately, we need to answer a key question: do we understand the relationship between our dirty map $\widehat{\x}$ and the input sky model from which we simulated visibilities? It is not important how much our dirty maps look like the sky itself. We just want to make sure that we keep track of everything the instrument and our mapmaking algorithm has done to the data so we can take it into account properly when start estimating power spectra.

We therefore need an input sky model for two reasons. First, we need to be able to use Equation \eqref{eq:AnalyticVisibility} to compute visibilities and thus $\widehat{\x}$. Next, we also want to compute the matrix of point spread functions $\PSF$ corresponding to the same set of observations and multiply it by our true sky model $\x$. The error metric we use therefore is
\beq
\varepsilon = \frac{\left| \widehat{\x}_\text{exact} - \widehat{\x}_\text{approx} \right|}{\left|\widehat{\x}_\text{exact} \right|}. \label{eq:errorMetric}
\eeq 
To be clear, this does not measure the difference between our dirty map and the true sky. It is merely a measure of the discrepancy between what the instrument and our mapmaking routine did to the sky in order to form the dirty map ($\widehat{\x}_\text{exact}$) and what we think we know about those effects ($\widehat{\x}_\text{approx}$) when we write down $\mean$ and $\C$.
  
One advantage to this metric is that it is often relatively easy to calculate $\widehat{\x}_\text{exact}$, at least up to $\D$ which we can factor out of the numerator of Equation \eqref{eq:errorMetric}, compared to calculating $\PSF$. That is because calculating $\A^\dagger \N^{-1} \y$ is as computationally difficult as calculating a single row of $\PSF$. In the following sections, we will be examining ways of computing $\PSF$ faster. Sometimes (e.g. in Sections \ref{sec:faceting} and \ref{sec:PSFfitting}) that means an approximate $\PSF$ but an exact $\widehat{\x}$, in which case $\widehat{\x}_\text{approx} = \PSF_\text{approx} \x$. Other times (e.g. in Section \ref{sec:snapshots}) that means a method for computing $\widehat{\x}$ that also makes $\PSF$ easier to compute. In that case, Equation \eqref{eq:errorMetric} compares the approximate method for computing $\widehat{\x}$ with the exact one. 

We have chosen a sky model with two components: 1) bright point sources and 2) diffuse emission from our Galaxy and other dim, confusion-limited galaxies. Since each frequency is measured and analyzed independently (meaning that $\A$ is sparse and can be written compactly in blocks), we will perform all the simulations at a representative  frequency of 150\,MHz. While the simulations properly weight visibilities based on how many times each unique baseline was measured, we do not include any noise in our calculation of the quantities in Equation \eqref{eq:errorMetric}. We also assume that all baselines at a given frequency have the same noise properties, though that assumption can be straightforwardly relaxed.

\subsubsection{Point Sources}   

\begin{figure}[] 
	\centering 
	\includegraphics[width=.5\textwidth]{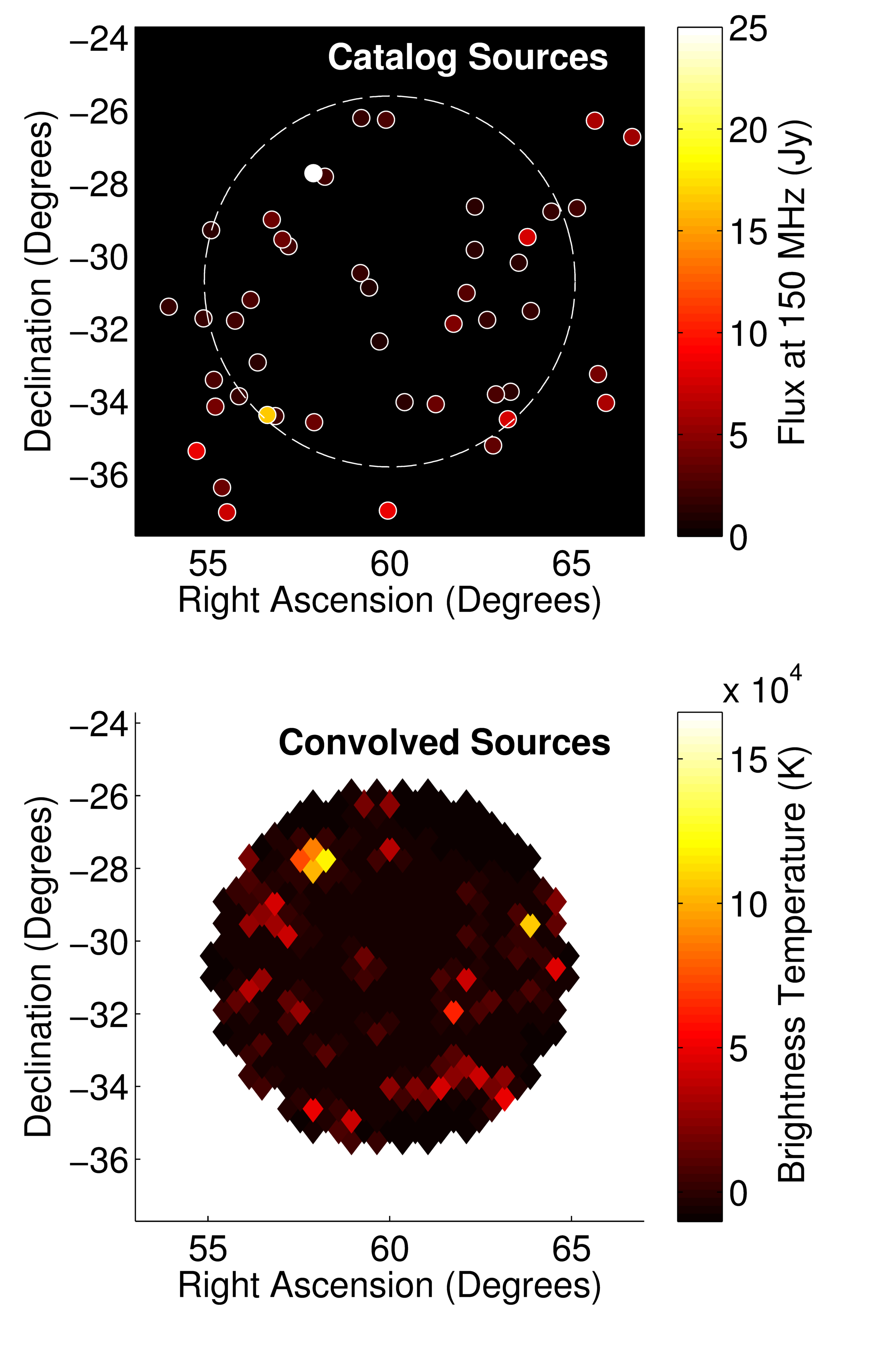}
	\caption{To test our mapmaking method and our approximate techniques for making it much faster, we need a fiducial sky model. One component of that model is bright point sources, which are taken from the MWA Commisioning Survey Catalog \cite{MWACS}. In the top panel, we show the spatial distribution and intrinsic flux of all point sources whose primary-beam-weighted fluxes are above 1\,Jy. In the bottom panel, we show $\widehat{\x}=\PSF\x$, the PSF-convolved and discretized dirty map with HEALPix $N_\text{side}=128$. Since the point spread functions are computed at the locations corresponding to each point source, the bottom panel is exact.}
	\label{fig:PointSources}
\end{figure}

Our sky model includes bright point sources above 1\,Jy with specified positions, fluxes, and spectral indices. These are taken from the MWA Commissioning Survey Catalog \cite{MWACS}, which is complete to below 1\,Jy for a large fraction of the sky. The included spectral indices are used to extrapolate their fluxes at 150\,MHz down from the survey frequency of 180\,MHz. For the calculation of visibilities using Equation \eqref{eq:AnalyticVisibility}, they are treated as true point sources with Dirac delta function spatial extent. In Figure \ref{fig:PointSources}, we show a representative sample of those point sources and what they look like in the dirty map, $\widehat{\x}$.

The sky model for point sources is completely independent of our pixelization. Since we know the location of all the point sources, we can think of $\x$ as having a discretized component covering the whole sky in pixels---which we will use for analyzing diffuse emission---and a set of Dirac delta function fluxes at the positions of the point sources. The sky model for point sources is completely independent of our pixelization. This is completely compatible with the definition of our pixelization in Section \ref{sec:interferometry}, it is just that some pixels have finite area and some have infinitesimal area. It is the pixels with finite volume that we care about for 21\,cm power spectrum estimation, but the infinitesimal ``pixels" matter for foreground subtraction. Likewise, $\PSF$ has two blocks: one that maps pixels on the true sky to pixels on the dirty map and one that maps points on the true sky to pixels on the dirty map.

\subsubsection{Diffuse Emission}

In the case of point sources, we might hope to use precise locations on the sky to refine our models of $\mean$ and $\C$ and do a better job of separating foregrounds from the 21\,cm signal. That is simply not possible with diffuse synchrotron emission from our Galaxy or with the confusion-limited emission from  relatively dim radio galaxies. Fundamentally, our best guess at that emission and its statistics will have to be discretized and pixelized. Uncertainty about how many confusion-limited point sources appear in a single pixel introduces shot noise, which can be modeled \cite{LT11,DillonFast}. 

In this work, we are interested in errors caused by assumptions and approximations in our mapmaking routine whose effects are not taken into account when estimating power spectra. In order to write down a vector $\x$ that we can use to compute $\widehat{\x}$ and thus $\varepsilon$ with Equation \eqref{eq:errorMetric}, we can either treat the emission as constant in the pixel or we can treat the emission as a ``point source" at the center of each pixel. For computational simplicity, we choose the latter. With relatively small pixels, there is no practical difference between the two. Since we are concerned about translating our models for foreground residuals in the true sky into models in the dirty map, the pixelization here is not an approximation so much as a consequence of the discretized models for foreground residuals we need for power spectrum estimation. It is possible to construct $\PSF$ to have different angular resolutions of $\x$ and $\widehat{\x}$, if one would like to incorporate a high-resolution diffuse foreground covariance model. The more information we can incorporate about the foregrounds, the smaller our uncertainties get and the better foreground subtraction works.

\begin{figure}[] 
	\centering 
	\includegraphics[width=.5\textwidth]{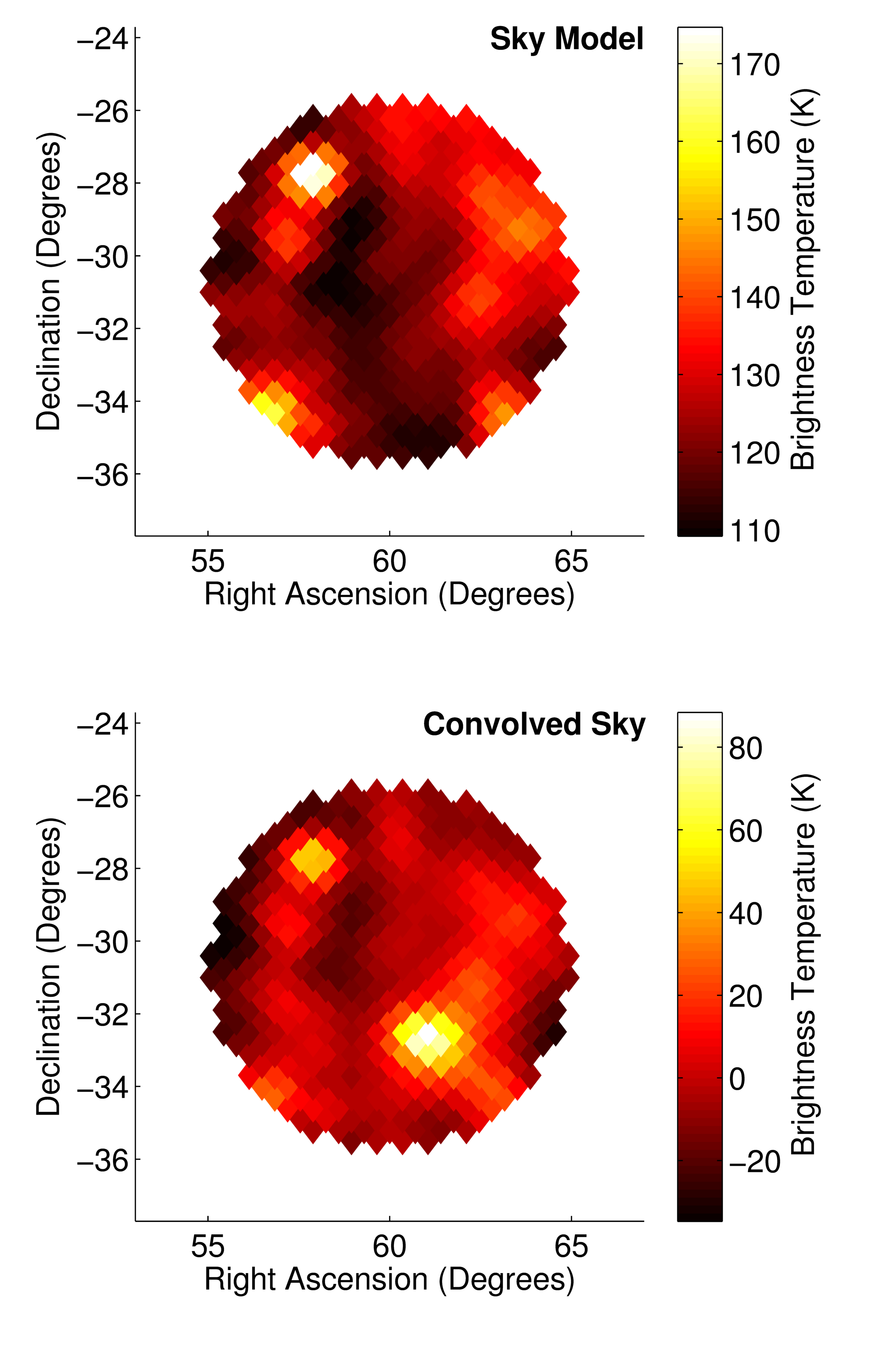}
	\caption{The sky model we use to evaluate our mapmaking algorithm and the accuracy of the approximations we make also includes diffuse emission from our Galaxy and faint radio galaxies. For our model of diffuse emission, we use the Global Sky Model of \cite{GSM}. In the top panel, we show a small part of our model for the true diffuse emission. Since we are not trying to model fine spatial information or the precise locations of point sources with our diffuse models, we pixelize the emission identically to the pixelization of our dirty map. In the bottom panel, we show that dirty map. It looks fairly different from the true sky, largely because of the appearance of a side lobe from a bright object outside the field. This occurs because the $\PSF$ maps a very large region of the sky to a small one shown here. The effects of faceting and side lobes will be explored further in Section \ref{sec:faceting}.}
	\label{fig:GSM}
\end{figure}

We use the popular HEALPix software package \cite{HEALPIX} for discretizing the celestial sphere into regularly spaced, equal-area pixels. As a model for the emission itself, we use the Global Sky Model of \citet{GSM} (see Figure \ref{fig:GSM}). The precise model we choose for this work matters only insofar as it is relatively realistic and representative of the true sky. That said, building good foreground models is an important ongoing endeavor relevant to power spectrum estimation and foreground subtraction \cite{ChrisMWA,PoberWedge,JacobsFluxScale,InitialLOFAR1,InitialLOFAR2}.
 
\subsection{Computational Challenges of Mapmaking} \label{sec:challenges}

We already alluded to the fact that we need to investigate various simplifications and approximations to make the calculation of $\widehat{\x}$ and $\PSF$ tractable. Let us take the time to see exactly where the problem lies.

Consider the matrix $\A$ where $\y = \A \x + \n$. $\A$ maps a discretized sky into time-ordered data. If we want to slightly over-resolve the sky with HERA, we might choose a HEALPix map with $N_\text{side} = 256$, which gives an angular resolution of about $0.2^\circ$. That is almost $10^6$ pixels at each of about 1000 different frequencies (assuming 100\,kHz resolution and 100\,MHz of simultaneous bandwidth). If we measure all our visibilities every two seconds for 1000 total hours at all 1000 frequencies, that is $10^{14}$ visibilities, so naively, $\A$ is a $10^{14} \times 10^9$ matrix. That is a problem.

Of course, there are many standard simplifications. Each frequency is treated completely independently during mapmaking, so we can treat $\A$ as either block diagonal or as a family of 1000 much smaller matrices, $\A(f)$. Redundant baselines measure the same sky, so their visibilities can be combined together, reducing both instrumental noise and the number of visibilities by a factor of almost 100 in the case of HERA. Getting 1000 hours of nighttime observation takes about 100 days, so we can LST-bin, reducing both noise variance and data volume by another two orders of magnitude. Since each time-step is independent of all others, we can further break $\A$ into about 10,000 pieces for each integration.  

We still have $10^7$ different $\A$ matrices, each $10^3 \times 10^6$. This size is challenging but acceptable for either simulating visibilities or calculating $\A^\dagger \N^{-1} \y$. However, it is simply too big for the calculation of $\PSF$, which would require the computationally infeasible task of multiplying together two matrices of this size $10^7$ times, each multiplication taking roughly $10^{15}$ operations. In the following sections, we will look at ways of reducing the number of $\A(f)$ matrices and making each $\A(f)$ smaller, especially during the calculation of $\PSF$.

\subsection{Faceting and First Mapmaking Results} \label{sec:faceting}

The matrix of point spread functions $\PSF$ is defined by the relation $\langle\widehat{\x} \rangle = \PSF \x$. It can be thought of as a transformation from one pixelized real space---that of the true sky---to another---that of the dirty map. For even a modest angular resolution, that is an enormous matrix. Do we really need to know the relationship between every pixel in the sky and every pixel in the dirty map? 

\subsubsection{Why We Facet} 

\begin{figure}[]  
	\centering 
	\includegraphics[width=.48\textwidth]{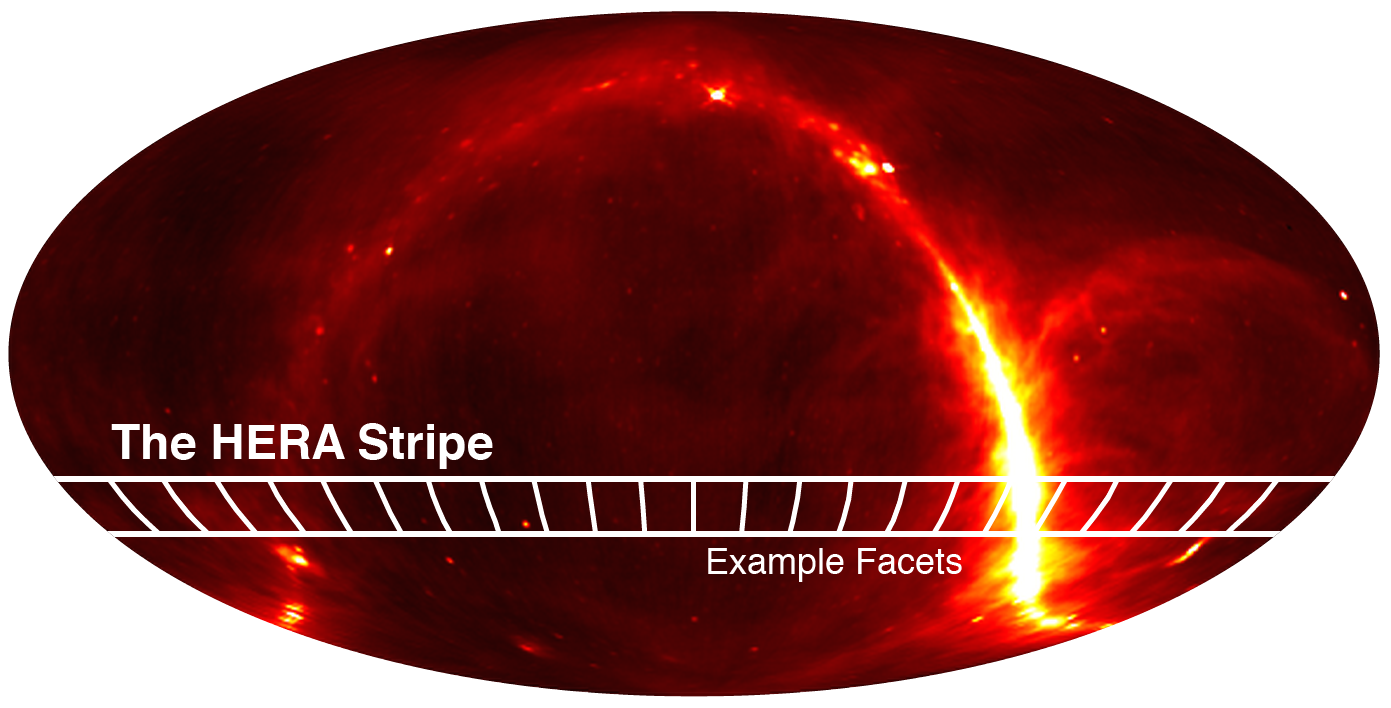}
	\caption{The faceted approach we use to speed up optimal mapmaking and power spectrum estimation will be especially useful for HERA because it is limited to only observe an approximately $10^\circ$ stripe of constant declination, centered on the array's latitude of approximately $-30.7^\circ$. It is fairly natural to split up the observation into roughly $10^\circ\times10^\circ$ facets, each analyzed separately. This makes $\PSF$ much easier to compute and lets us use the flat-sky approxmation, a requirement for implementing the power spectrum methods of \citet{DillonFast}. Very little cosmological information is lost in this process; only the longest spatial modes are thrown out and they should be dominated by galactic emission.}
	\label{fig:HERA_Stripe}
\end{figure}

Breaking up the field of view into a number of smaller facets is a standard technique in radio astronomy, especially when one wants to minimize the effects of noncoplanar baselines \cite{CornwellWProj}. For purposes of 21\,cm cosmology, there are two good reasons to consider relatively small regions of the sky one at a time. The first is HERA's observing strategy. Because it statically points at the zenith, HERA scans a fixed stripe in declination about $10^\circ$ degrees wide.  It seems reasonable that we can analyze parts of the stripe independently, making maps and computing power spectra for each small facet. In Figure \ref{fig:HERA_Stripe}, we show an example of what that faceting might look like.

The only significant disadvantage to faceting is that we lose the ability to measure modes in the power spectrum with wavelengths perpendicular to the line of sight that are larger than the facet. Doing so properly and with precisely quantified error properties would require calculating covariance between facets, which is effectively the same as not faceting at all. This is not such a great hardship. Due to the survey geometry, only the long modes oriented along the HERA stripe could have been measured at all. They are longer than the shortest baseline, meaning that they can only be sampled after considerable sky rotation.  The same $|\mathbf{k}|$ modes can be also be accessed along the line of sight, except those at very low spectral wave-numbers, which are bound to be foreground dominated.

The other major upside to faceting is that, if we want to use the fast power spectrum techniques developed in \cite{DillonFast}, we need to take our maps and chop them up into facets anyway. That is because any fast algorithm that takes advantage of the fast Fourier transform (e.g. that in \cite{DillonFast}) and translational invariance relies on rectilinear data cubes, which is only an accurate approximation for small fields where the flat-sky approximation holds. Happily, that rough size is also about $10^\circ$. For other instruments, the choice of facet size is less obvious and depends on the computational demands of both mapmaking and power spectrum estimation. Bigger facets preserve more information, but they can be more computationally expensive than they are cosmologically useful. The exact right choice for other interferometers is a matter for future work.

\subsubsection{Faceted Mapmaking Method And Results}

So, instead of using $\D\A^\dagger \N^{-1} \y$ to calculate $\widehat{\x}$, we instead redefine $\widehat{\x}$ using
\beq
\widehat{\x} = \D \K_\text{facet}\A^\dagger \N^{-1} \y,
\eeq
where $\K_\text{facet}$ maps the full sky to a small portion of the sky, thus making $\PSF$ asymmetric. Doing this for every facet basically amounts to only mapping the parts of the sky that are ever near the center of the primary beam. This provides a computational simplification by a factor of $4\pi / (\Omega_\text{facet} N_\text{facets})$, which for HERA is about an or order of magnitude.  An instrument that can see the whole sky would see no computational benefit just from breaking the sky in facets.

The real computationally limiting step is the calculation of $\PSF$. Since we are only interested in the dirty map of a facet, we care only about source flux that could have contributed to that dirty map. That means that we can truncate each point spread function some distance from the facet center. Flux outside that truncation radius is assumed not to contribute significantly. In other words, 
\beq
\PSF = \D \K_\text{facet} \A^\dagger \N^{-1} \A \K_\text{PSF}^\trans
\eeq
where $\K_\text{PSF}$ is the same as $\K_\text{facet}$ except that it cuts off at some larger radius than the facet size. We get to choose exactly what radius we want to assume that no outside flux contributes to the facet. This is a completely tunable approximation and it becomes exact in the limit that that radius encompasses the whole sky. 

Therefore, instead of mapping the whole sky to the whole sky, the matrix of point spread functions now maps some moderate portion of the sky to a somewhat smaller part of the sky. Since $\N$ is diagonal, both the time it takes to calculate $\PSF$ and the memory it takes to store it are reduced by very large factor. If the truncation region is 4 times the $10^\circ$ facet size, for example, then that savings is a factor of about $10^4$.

This new definition of $\widehat{\x}$ means that $\D$ is now a much, much smaller matrix---it has only as many elements as there are pixels in the facet. And since we are only interested in the correlation between pixels in the map, the noise covariance is now
\beq
\C^N = \PSF \K_\text{facet}^\trans \D^\trans,
\eeq
which is much smaller and still quite simple.

\begin{figure*}[] 
	\centering 
	\includegraphics[width=1\textwidth]{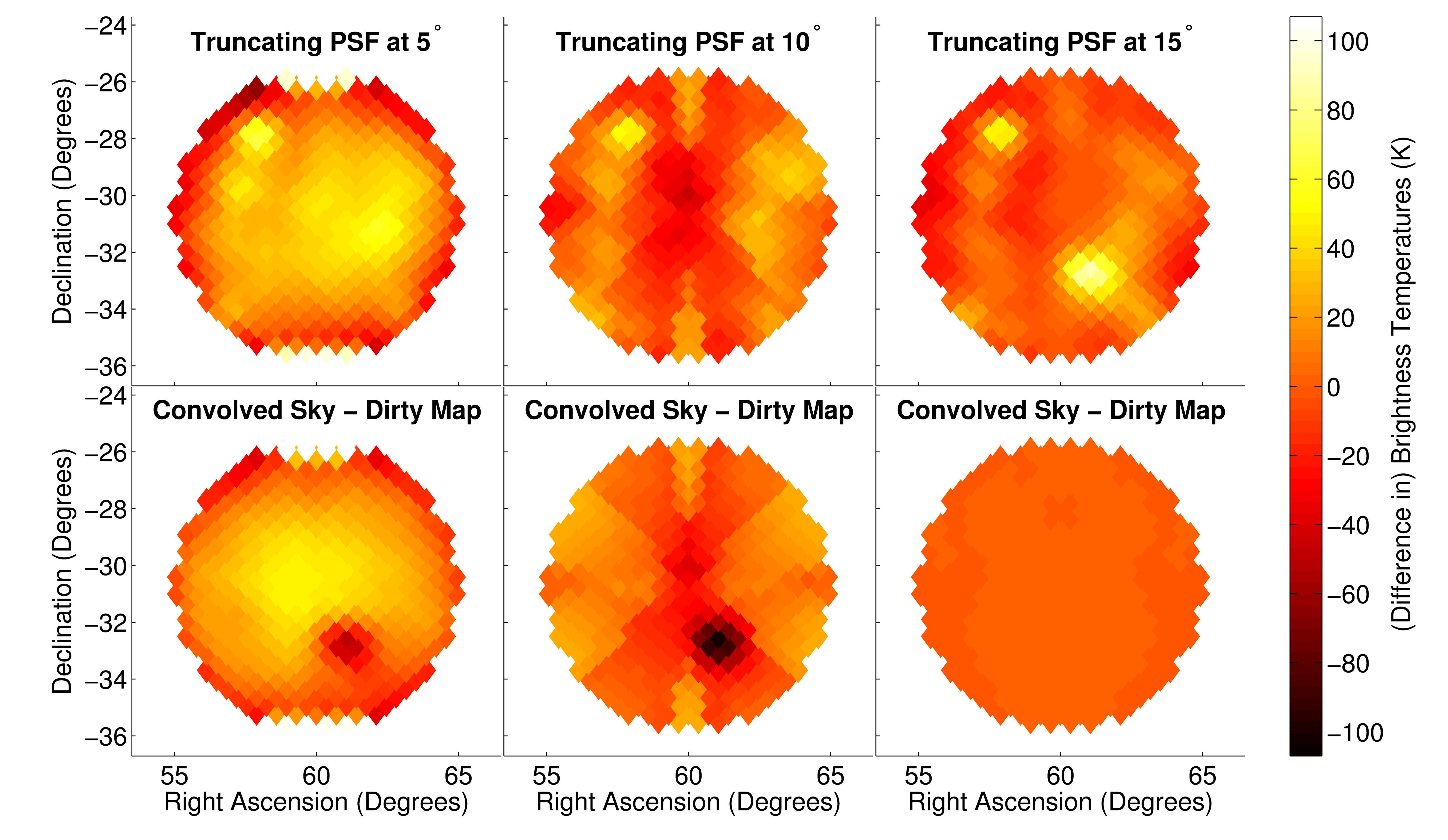}
	\caption{In order to accurately reproduce dirty maps, we must include in our $\PSF$ matrix the effect flux from outside the facet that appears in the side lobes of off-facet sources. Here we demonstrate that effect by looking at how the approximate PSF-convolved sky, $\PSF_\text{approx}\x$, evolves as we expand the distance from the center of the facet at which the point spread function is approximated to not contribute. In the top row, we plot $\PSF_\text{approx}\x$ while on the bottom row we plot $\PSF_\text{approx}\x - \widehat{\x}_\text{exact}$. ($\PSF_\text{exact}\x = \widehat{\x}_\text{exact}$ is shown in the bottom panel of Figure \ref{fig:GSM}.) Since the visibilites that go into computing $\widehat{\x}$ derive from a full-sky calculation, side lobes are automatically included. The bright spot we see on the top right panel, which appears as a dark spot on the bottom left and bottom middle panels, is a prominent side lobe from a very bright source outside the facet, but within $15^\circ$ of the facet center. This explains what we saw in Figure \ref{fig:GSM} and the dramatic improvement in the error we see in the right-hand panels.}
	\label{fig:PSF_Progress}
\end{figure*}

We illustrate the effect of the PSF truncation radius in Figure \ref{fig:PSF_Progress}, showing the large impact that increasing the truncation radius has on our calculations of $\widehat{\x}_\text{approx} = \PSF_\text{approx} \x$ and therefore of $\varepsilon$. We find that once the PSF includes both the central peak of the synthesized beam and the first major side lobes, the convergence of $\widehat{\x}_\text{approx}$ to $\widehat{\x}_\text{exact}$ is very quick.

\begin{figure}[] 
	\centering 
	\includegraphics[width=.5\textwidth]{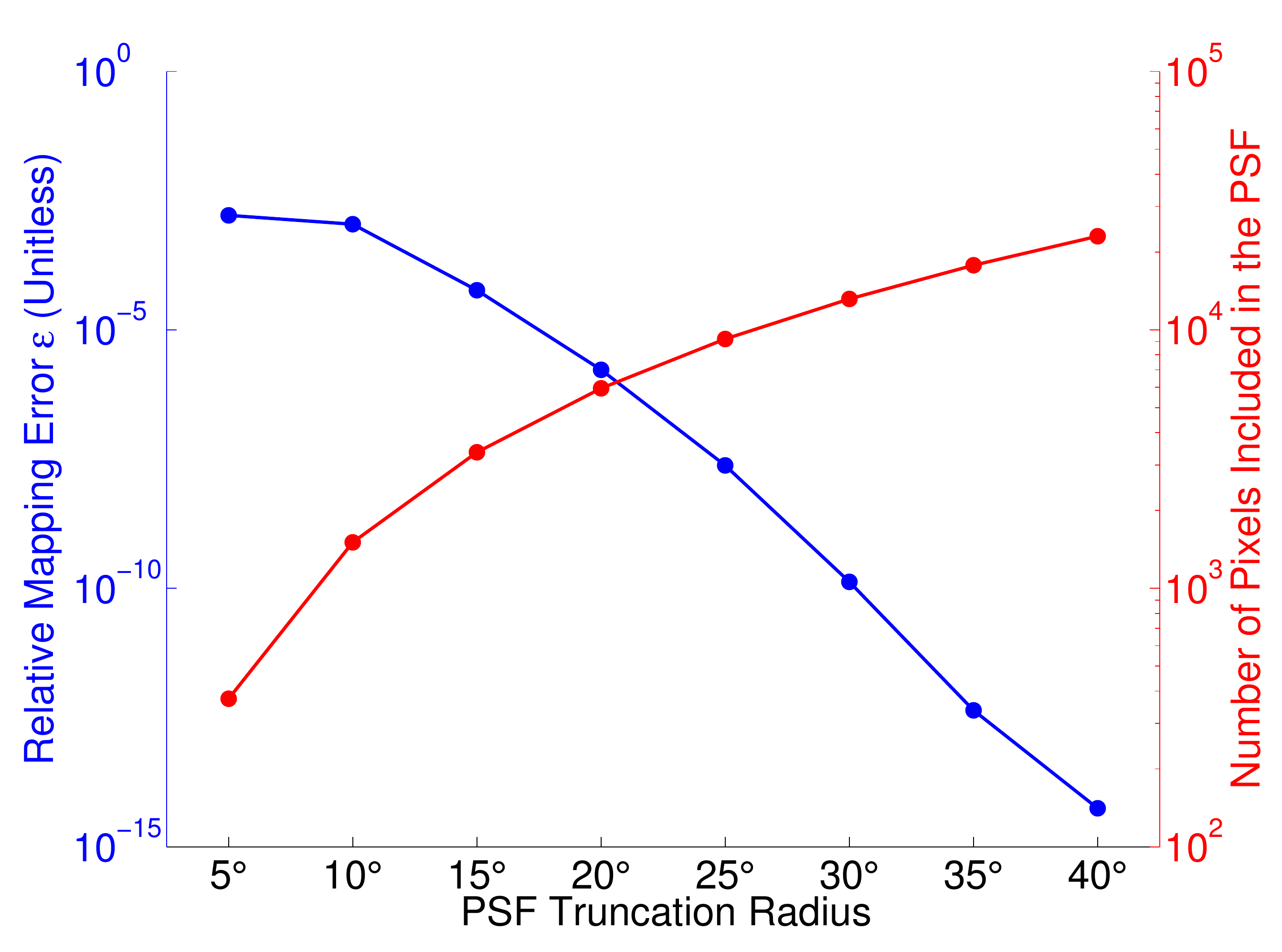}
	\caption{The error introduced by the approximation that the PSF can be truncated past a certain distance from the facet center gets very small very quickly. Here we show how both that error, which we define in Equation \eqref{eq:errorMetric}, and the number of pixels in each point spread function depend on the truncation radius. The number of pixels, and thus the computational difficulty of computing the matrix of point spread functions, $\PSF$, scales as the truncation radius squared---there are simply more pixel values to calculate. In general, the approximation works because the point spread functions are relatively compact. HERA's design is especially helpful here with its dense grid of baselines and its relatively small primary beam. Other arrays may need larger truncation radii to acheive the same accuracy.}
	\label{fig:PSF_Size_Test}
\end{figure}

We further tested the expected convergence of the algorithm for a fixed facet size and variable $\K_\text{PSF}$ using the sky model from Section \ref{sec:skymodel}. Our results, which we show in Figure \ref{fig:PSF_Size_Test}, again demonstrate that the PSF truncation radius does not need to be much larger than the facet, if the facet is comparable in size to the primary beam. The exact level of error introduced by faceting will, in general, depend upon the compactness of both the primary and synthesized beams. The approximation that the point spread function is Gaussian might make the plotted relative error a bit optimistic, though the side lobes in the real HERA primary beam are quite small.

In summary, faceting allows us to decrease the time it takes to calculate the $\PSF$ and the memory required to store it by a factor of $(4\pi)^2 / (\Omega_\text{facet} \Omega_\text{PSF})$, where $\Omega_\text{PSF}$ is the angular size of the region left by $\K_\text{PSF}$. In the case of HERA, that works out to about 10,000 times faster and smaller.

\subsubsection{Mitigating Nonredundancy} \label{sec:nonredundant}

Making maps in facets also has one extra advantage useful in addressing a common complication presented by real-world arrays. If we assume in our analysis that every baseline of a given designed separation actually has that separation, we will be ignoring errors that can be a decent fraction of a wavelength. And though HERA is a zenith-pointed array for which noncoplanar effects are small, they are not zero and can be quite large for other instruments like the MWA. Noncoplanarity creates nonredundancy. 

However, as long as we know precise positions of all of our antennas (which is far easier than making the array perfectly redundant) we can use the fact that we are only mapping a single facet at a time to reduce those phase errors near the center of our map. We can think of each baseline corresponding to some unique baseline $\base$ as
\beq
\base_m = \base + \Delta \base_m,
\eeq
where the residuals are caused by inexact antenna placement. That means that Equation \eqref{eq:VisibilitySum} becomes 
\begin{align}
V(\base_m,\nu_n) \approx \sum_k &\Delta\Omega \frac{2 k_B \nu_n^2}{c^2} x_k(\nu_n) B(\rhat_k,\nu_n) \times \nonumber \\
& \exp\left[-2 \pi i  \frac{\nu_n}{c} \left(\base + \Delta \base_m\right) \cdot \rhat_k \right]. \label{eq:nonredundant}
\end{align}
We need the right-hant side of this equation to be the same for all $\base_m$ corresponding to the unique baseline $\base$, otherwise we lose the redundancy bonus we discussed in Section \ref{sec:challenges}. 

We can achieve this approximately for small $\Delta \base_m$ because our facets are relatively small. Let us define $\Delta \rhat_k \equiv \rhat_k - \rhat_0$ where $\rhat_0$ points to the center of the facet and $\Delta\rhat_k$ is generally not a unit vector. We can expand the exponent of Equation \eqref{eq:nonredundant} as
\begin{align}
&\left(\base + \Delta \base_m\right)\cdot \left( \rhat_0 + \Delta\rhat_k \right)  \nonumber \\
= & \mbox{ } \base\cdot \rhat_0 + \base\cdot\Delta\rhat_k + \Delta\base_m\cdot\rhat_0 + \Delta\base_m \cdot \Delta\rhat_k.
\end{align}
The first two terms in the expansion are $\base\cdot\rhat_k$ and normally appear in $\A$. The last term, which second order in this expansion, is approximated to be zero. Even if $\base \cdot \rhat_0$ is small, the last term is in general much smaller than the second term. We can, however, correct for the middle term by multiplying both sides of Equation \eqref{eq:nonredundant} by a constant phase factor, since
\beq
V(\base, \nu_n) \approx \exp\left[2 \pi i  \frac{\nu_n}{c} \Delta \base_m \cdot \rhat_0 \right] V(\base_m, \nu_n).
\eeq
As was our goal, the $\PSF$ matrix that results from taking the above equation to be exactly true is the same as if we had not had any antenna placement errors or noncoplanarity. Rephasing lets us mitigate the effect of known errors without having to calculate a vastly more complicated $\PSF$, which treats all baselines completely independently, even if they are supposed to be redundant.

Effectively, our approximate correction cancels out the phase error at the exact center of the facet and thus minimizes its effect throughout the facet. For example, for $10^\circ$ facets at 150\,MHz, a 4\,cm antenna placement error (roughly the level seen in \cite{MITEoR}) leaves only a $0.63^\circ$ phase error in the visibility after rephasing. The error might be a bit worse when calculating the parts of $\PSF$ near the truncation radius. For very large fields, as \cite{CornwellWProj} addressed, this becomes a bigger problem and we may need to break each set of baselines that was supposed to be redundant into a few groups, each closer to exactly redundant, and treat each group separately. The exact effect on the accuracy of the dirty maps from this small correction is left to future work when the exact antenna placement of HERA or a similar array is known. 

\subsection{Grouping Visibilities into Snapshots} \label{sec:snapshots}

Standard interferometric mapmaking techniques accumulate visibilities in the $uv$-plane via sky rotation and thereby combine minutes or even hours of visibilities together \cite{Schwab1984,Cotton2004,Cotton2005,Bhatnagar2008,Carozzi2009,Mitchell2008,bernardi,Bart,Smirnov2011,Li2011,FHD,WSCLEAN,TrottObservingModes}. We would like to find a way of reducing the number of rows in $\A$ for the purpose of calculating $\PSF$ by grouping integrations into ``snapshots" that are each analyzed as a single timestep when we calculate $\PSF$. How can we average together multiple visibilities over a range of times while approximating the $\PSF$ as having been calculated at only the middle timestep of each snapshot? 

Once again, we can use our freedom to rephase both the visibilities and the $\A$ matrix as we did in Section \ref{sec:nonredundant}. The idea is to try to remove, as much as possible, the effect of sky rotation from the visibilities. Consider again Equation \eqref{eq:AnalyticVisibility}, now with explicit time dependence:
\begin{align}
V(\base,\nu,t) = \int & B\left(\rhat,\nu \right)
I(\rhat,\nu,t) \times \nonumber \\
&\exp \left[-2 \pi i  \frac{\nu}{c} \base \cdot \rhat \right] d\Omega.
\end{align}
While the sky rotates, the primary beam is fixed relative to the ground. 

By contrast, let us consider a new reference frame with angle vector $\rhat'$, which rotates with the sky:
\begin{align}
V(\base,\nu,t) = \int&  B\left(\rhat',\nu,t \right)
I(\rhat',\nu) \times \nonumber \\
&\exp\left[-2 \pi i  \frac{\nu}{c} \base(t) \cdot \rhat' \right] d\Omega'.
\end{align}
Now the beam and the baseline vector have picked up an explicit time dependence while the sky has lost its time dependence. Let us assume that the primary beam is varying very slowly spatially---generally a good assumption since the primary beam is much larger than the spatial scales probed by most baselines. 

Let us think of $V(\base,\nu,t)$ as the visibility measured for the middle integration of a snapshot. A visibility measured a bit later during that snapshot would look like
\begin{align}
V(\base,\nu,t+\Delta t) \approx& \int d\Omega' B\left(\rhat',\nu,t \right)
I(\rhat',\nu) \times \nonumber \\
&\exp\left[-2 \pi i  \frac{\nu}{c} (\base(t)+\Delta \base) \cdot \rhat' \right],
\end{align}
where $\Delta \base$ is the difference between $\base(t+\Delta t)$ and $\base(t)$ in the primed coordinate system. The dot product is basis independent, so 
\beq
(\base(t)+\Delta \base) \cdot \rhat' = \base\cdot \left(\rhat + \Delta \rhat(\rhat) \right),
\eeq
where the right-hand side is back in the frame that is stationary relative to the Earth. $\Delta \rhat(\rhat)$, which is not a unit vector, is the amount of sky rotation between times $t$ and $t+\Delta t$. It is approximately constant across the facet for fairly short snapshots and moderately sized facets, meaning that we can pull it out of the integral. We can therefore undo much of the effect of sky rotation using the approximation that
\beq
V(\base,\nu,t+\Delta t) \approx e^{i \Delta\phi} V(\base,\nu,t)
\eeq
where
\beq
\Delta\phi \equiv -2 \pi  \frac{\nu}{c}  \base \cdot (\rhat_0(t+\Delta t) - \rhat_0(t))  
\eeq
and where again, $\rhat_0(t)$ points to the facet center. 

\begin{figure*}[] 
	\centering 
	\includegraphics[width=1\textwidth]{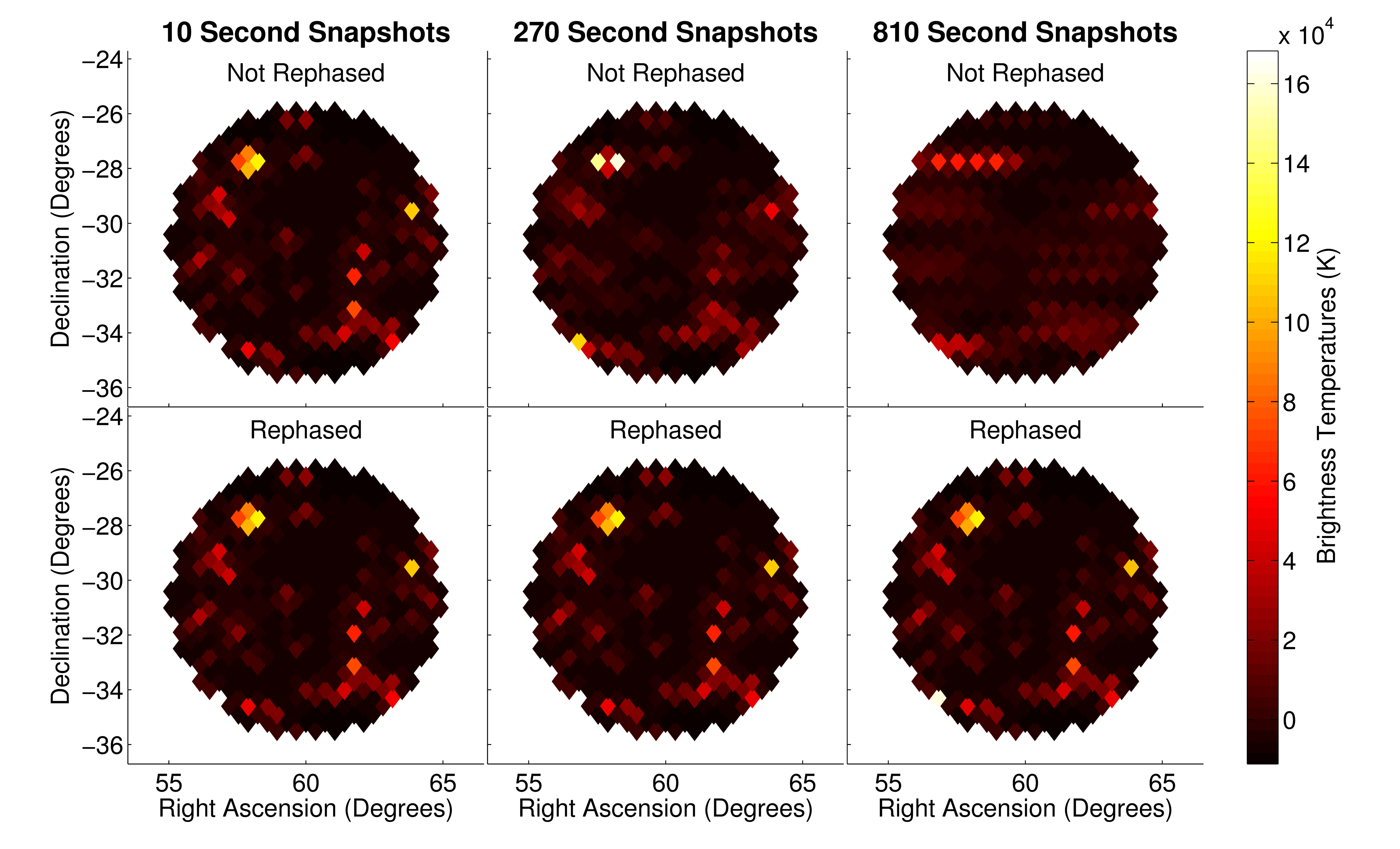}
	\caption{One way to make the calculation of the matrix of point spread functions, $\PSF$, faster is to combine many consecutive integrations together into snapshots. When we compute $\PSF$, we effectively assume that all the associated visibilities we have grouped into one snapshot were taken exactly at the snapshot's middle time. Usually, this is a poor approximation. As we can see from the top row, where we have simply added together 10 second integrations to snapshots of increasing length, we are effectively spreading out sources in right ascension as the sky rotates overhead. However, if we use our freedom to rephase visibilities individually, we can dramatically reduce the error associated with forming snapshots. For example, the bottom right panel only exhibits error on the order of a few percent compared to the exact single-integration dirty maps in the left-hand panels.  The result is related to the traditional radio astronomy technique of ``fringe stopping."}
	\label{fig:Snapshot_Progress}
\end{figure*}

We can therefore add together many visibilities taken at different times and approximately treat them as if there were all taken at the middle integration in the snapshot by rephasing them. This is very similar to the ``fringe-stopping" technique from traditional radio astronomy, which seeks to counteract the effect of the rotation of the earth at the location of a source \cite{ThompsonMoranSwenson}. As we saw in Section \ref{sec:nonredundant}, the effect of rephasing visibilities cancels out in $\PSF$, since the extra term in $\A$ gets canceled out in $\A^\dagger.$ That is why we only have to perform the calculation of $\PSF$ once per snapshot rather than once per integration. We show in Figure \ref{fig:Snapshot_Progress} a marked improvement, especially in the case of long snapshots, between naively adding together visibilities as if the sky were not rotating overhead and adding together rephased visibilities.

\begin{figure}[] 
	\centering 
	\includegraphics[width=.5\textwidth]{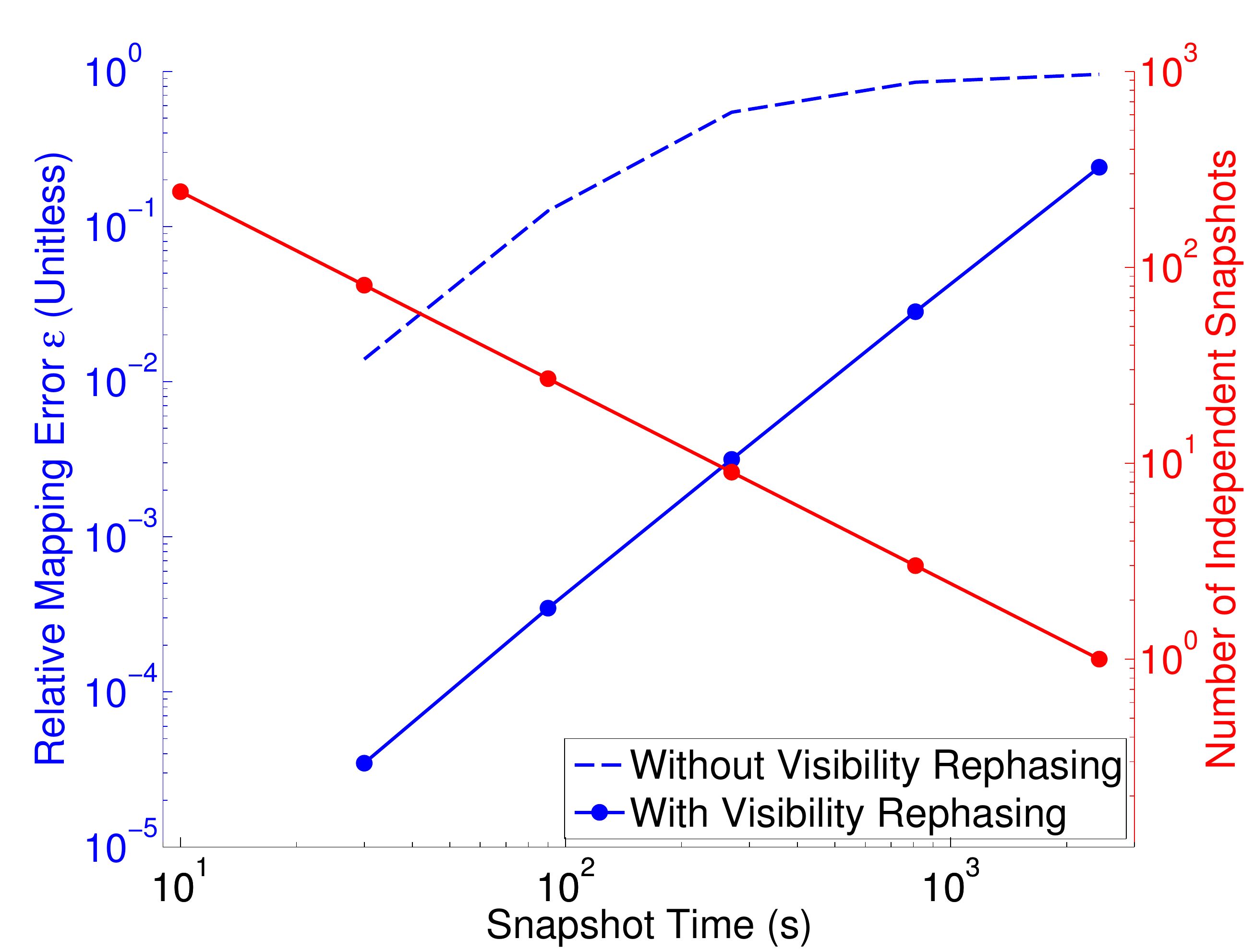}
	\caption{The error introduced by approximating the observation as having taken place at at only a few discrete times, many seconds or minutes apart, can be mitigated by appropriately rephasing visibilities before combining them. Here we show quantitatively how the length of snapshots---all multiples of the 10 second integration time used in our simulation---introduces small errors. We calculate the relative error $\varepsilon$ between dirty maps calculated with a given integration time and those calculated exactly using only one integration per snapshot. We also show how the computational difficulty of calculating $\PSF$ is affected, since it scales linearly with the number of independent snapshots considered.}
	\label{fig:Snapshot_Time_Test}
\end{figure}

In Figure \ref{fig:Snapshot_Time_Test} we show quantitatively how the error increases as snapshots get longer. Here we care how these approximate dirty maps compare to the exact dirty maps made when each 10\,s integration is treated completely separately. We also found it important to rephase the visibilities to the exact middle of the snapshot, which creates a first-order cancellation that removes some of the error associated with this approximation.

Based on the results we show in Figure \ref{fig:Snapshot_Time_Test}, it is likely that we can cut another one to two orders of magnitude off the total number of operations we need to perform to calculate $\PSF$, making that calculation considerably easier. For a given accuracy goal, it is also possible to make the calculation of $\PSF$ even simpler by forming snapshots with different durations for baselines of different lengths, keeping $\Delta \phi$ small. 

\subsection{PSF Fitting} \label{sec:PSFfitting}

Now that we have found accurate and well-understood approximations that make computing $\PSF$ computationally feasible, we need to worry about multiplying a vector by $\PSF$. This is a necessary step in any power spectrum estimation scheme adapted from \cite{DillonFast}, since $\PSF$ appears in Equations \eqref{eq:CS}, \eqref{eq:CcommaAlpha}, \eqref{eq:CN}, and \eqref{eq:CFG}. In general, the number of operations in this calculation scales with the number of pixels in the facet, the number of pixels in the PSF, and the number of frequency channels, i.e. as $\BigO{N_\text{facet} N_\text{PSF} N_f}$. This is slower than we would like, so we will endeavor to show how it can be sped up.

If the point spread function were constant across the field---if it looked the same in the top and bottom rows of Figure \ref{fig:PSFs}---then the solution would be simple. We could calculate only one PSF and then use it to fill out all of $\PSF$. Then, if we approximate HEALPix pixelization as a regular grid---which is true in the flat-sky approximation---we can write $\PSF$ using Toeplitz matrices. A Toeplitz or ``constant-diagonal" matrix represents a translationally invariant relationship.\footnote{Toeplitz matrices have a number of nice properties, including the fact that an $N\times N$ Toeplitz matrix can be multiplied by a vector in $\BigO{N\log N}$ operations. This is because the translational invariance lets us use the fast Fourier transform. See \cite{Toeplitz} for a review of these matrices and their properties or \cite{DillonFast} for a previous application to 21\,cm cosmology of the same relevant properties.} A Toeplitz matrix $\T$ has the property that each element only depends on its distance from the diagonal of the matrix, or in other words that
\beq
T_{ii'} = t_{i-i'}. \label{eq:Toeplitz}
\eeq
We can imagine that, if any part of the PSF can be fully represented by its displacement from the facet center, then we can write $\PSF$ for each frequency and facet as a tensor product of two matrices, each describing translational invariance along one of the two principal axes of the HEALPix grid.\footnote{We define the axes by taking the center pixel and computing the linearly independent vector directions towards the nearest two pixels. It is not a problem that these two directions are not orthogonal---the FFT can be performed along nonorthogonal directions, as pointed out by \cite{FFTT2}.} If we index along those axes with $i$ and $j$ in the dirty map and $i'$ and $j'$ in the true sky, then for a single frequency the matrix of point spread functions can be written as
\beq
P_{ii'jj'} = t_{i-i'}s_{j-j'} 
\eeq
or as
\beq
\PSF = \mathbf{T} \otimes \mathbf{S}
\eeq
where $\mathbf{T}$ and $\mathbf{S}$ are Toeplitz matrices.

And yet we can easily see from Figure \ref{fig:PSFs} that point spread functions do not respect translational invariance. In the bottom row where the PSFs are displaced from the center of the facet, the side lobes nearer the edge of the primary beam are downweighted relative to those nearer the center. This is a consequence of optimal mapmaking, which downweights the contribution from regions of the sky that the telescope is less sensitive to. However, we expect that the physical effects that lead to a translationally varying PSF, like the primary beam and the projected array geometry, should change smoothly over the field. So while the PSF is translationally varying, perhaps its translational variation can be modeled with a small number of parameters.

If we calculate $\PSF$, the matrix of point spread functions that maps every pixel in some extended facet to every pixel on the facet of interest, we can model this translational variation by reorganizing $\PSF$. We have chosen our normalization $\D$ so that the specific point spread function mapping the sky onto a given pixel has a value of 1 at the center pixel of its main lobe.  But what about all the pixels displaced exactly pixel northeast from the center of the main lobe in all the PSFs? Or ten pixels? 

\begin{figure*}[] 
	\centering 
	\includegraphics[width=1\textwidth]{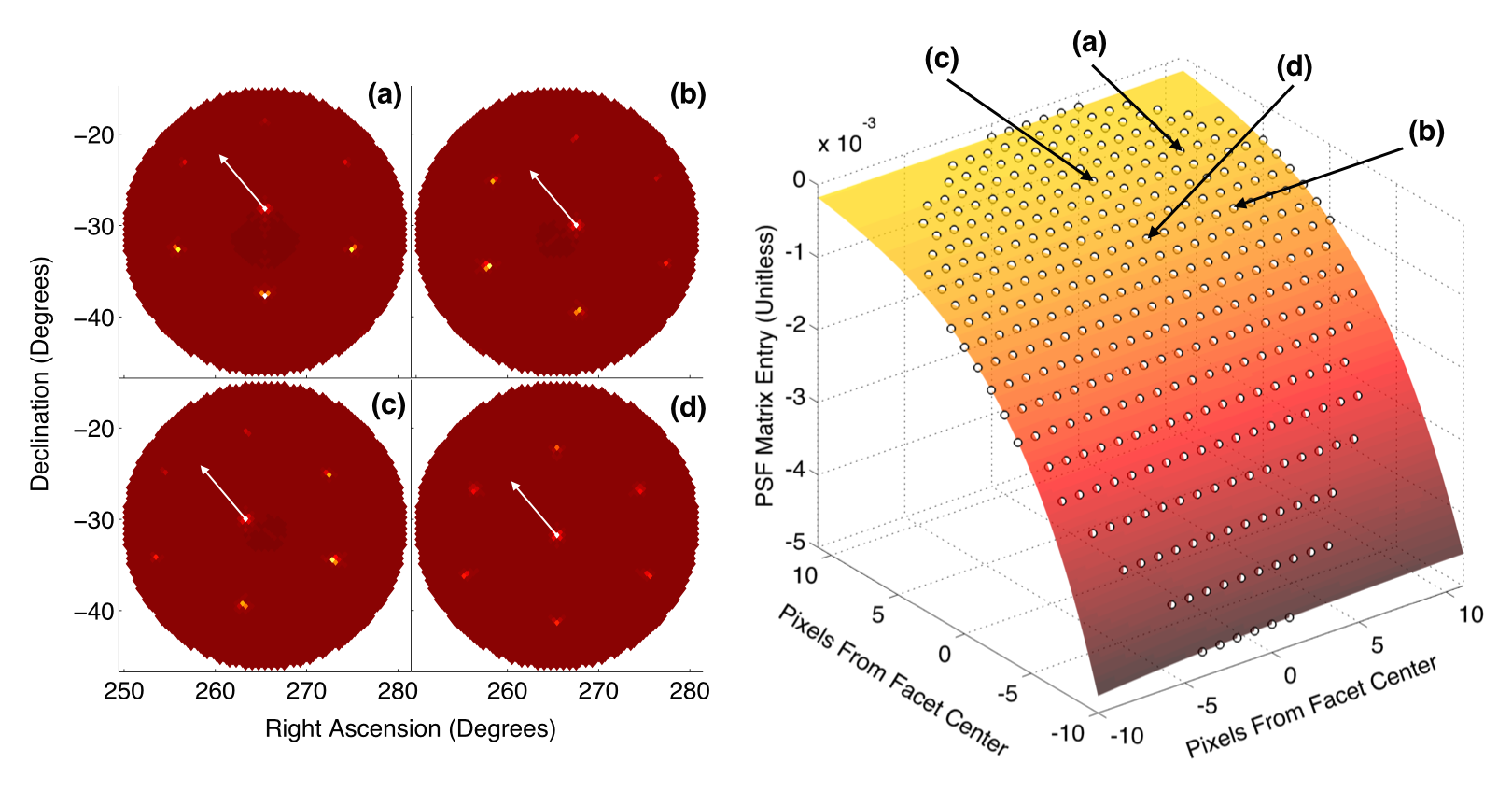}
	\caption{Though our point spread functions are not translationally invariant---a fact we saw clearly in Figure \ref{fig:PSFs}---their translational variation is fairly smooth and can be captured by a relatively low order polynomial. In this figure, we examine a typical example consisting of all the entries in $\PSF$ displaced exactly 15 pixels along one of the two principal axes of the pixelization from the center of the main lobe of the synthesized beam. This displacement is represented by the four identical white arrows on top of the point spread functions in the left-hand panel. All such entries in $\PSF$ (white circles in the right-hand planel) are plotted as a function of the displacement of the corresponding main lobe from the facet center. The points indicated by the white arrows in panels (a) through (d) are the same as the white circles indicated on the right hand plot. We then fit those points as a low-order 2D polynomial (in this case, as a quartic), which we plot as a colored plane cutting through them. The fit on the right hand side is merely one in a family of fits to each possible displacement vector from the main lobe of the PSF. Fitting the translational variation of the PSF in this way is potentially very useful, since a sparse representation of $\PSF$, the matrix of point spread functions, would allow us to quickly multiply it by a vector. Though this is not important for mapmaking, it is important for estimating power spectra from the dirty maps and mapping statistics produced by our method. }
	\label{fig:PSF_Fitting_Demo}
\end{figure*}

We expect these all to be similar, but also to vary slowly over the facet---though exactly how is not obvious \emph{a priori}. In the right-hand panel of Figure \ref{fig:PSF_Fitting_Demo} we plot the points on the PSFs displaced exactly 15 pixels along one of the two principal axes from the centers of their main lobes (illustrated by the left-hand panel). The $x$ and $y$ axes of the plot tell us which pixel a given PSF is centered on. As we expected, the variation over the facet is very smooth and is well approximated by a low-order polynomial. If we had instead plotted a displacement of 0, the right-hand panel would have been a perfectly flat plan of all ones because of the definition of $\D$.

How can we take advantage of the sparsity of information needed to describe $\PSF$ to write it as the sum of matrices that can be quickly multiplied by a vector? Let us first consider the simpler, 1D case. Instead of the translational invariance that leads to matrices of the form in Equation \eqref{eq:Toeplitz} where the main diagonal and all parallel off diagonals are constant, instead we model them all as polynomials:
\beq
P^\text{1D}_{ii'} = \sum_n t_{n,i-i'}(i+i')^n. \label{eq:polyToeplitz}
\eeq
This is a polynomial expansion in $(i+i')$, the distance along a diagonal, with coefficients $t_{n,i-i'}$ that make up a Toeplitz matrix. Again, primed indices tell us where on the true sky and unprimed indices tell us where in the faceted dirty map. The polynomial fit coefficients are a function of specific displacement of the main lobe of the PSF, hence the index $i-i'$. However, to fit all PSF values for the same displacement, we need to multiply those coefficients by the displacement from the center of the facet to the correct polynomial power. Our hope is that we can approximate $\PSF$ with a relatively low-order polynomial.

Expanding this out and cutting off the series after the second order in $n$, we get that
\beq
\PSF^\text{1D} \approx \T_0 + \J\T_1 + \T_1\J + \J^2\T_2 + 2\J\T_2\J + \T_2 \J^2
\eeq
where each $\T_n$ is a Toeplitz matrix and $\J$ is a diagonal matrix with integer indices centered on zero as its entries:
\beq 
\J \equiv  \text{diag}\left(...,-4, -3, -2, -1, 0, 1, 2, 3, 4, ... \right).
\eeq
Terms in the expansion that involve $(i')^n$ look like $\J^n$ to the right of $\T_n$, since they index into a vector multiplied by $\PSF^\text{1D}$ on the right, like the true pixelized sky. Likewise, terms that involve $i^n$ have a $\J^n$ matrix on the left.

In 2D, the situation is a bit more complicated. For clarity, let us treat $\PSF$ as a 4-indexed object, mapping two spatial dimensions to two other spatial dimensions. We approximate $\PSF$ as a polynomial sum of the form
\beq
P_{ii'jj'} = \sum_{n,m} t_{n,m,i-i',j-j'}(i+i')^n(j+j')^m.
\eeq
Now $\T_{n,m}$ is a ``block Toeplitz" matrix, essentially a Toeplitz matrix of Toeplitz matrices. Thankfully, multiplying by the matrix by a vector of size $N_\text{PSF}$ still only scales as $\BigO{N_\text{PSF} \log N_\text{PSF}}$ \cite{BlockToeplitz}. Expanding this to second order yields quite a few more terms:
\begin{align}
&\PSF \approx  \overbrace{\T_{0,0}}^{\text{$0^\text{th}$ Order}} + \nonumber \\
&\overbrace{\T_{1,0}(\J\otimes\Eye) + (\J\otimes\Eye) \T_{1,0} }^{\text{$1^\text{st}$ Order}} \nonumber +\\ 
&\T_{0,1} (\Eye\otimes\J) + (\Eye\otimes\J)\T_{0,1} +\nonumber \\
&\overbrace{\T_{2,0}(\J^2\otimes\Eye) + 2(\J\otimes\Eye)\T_{2,0}(\J\otimes\Eye) + (\J^2\otimes\Eye)\T_{2,0}}^{\text{$2^\text{nd}$ Order}} + \nonumber \\
&\T_{1,1}(\J\otimes\J) + (\J\otimes\Eye)\T_{1,1}(\Eye\otimes\J) + \nonumber \\
&(\Eye\otimes\J)\T_{1,1}(\J\otimes\Eye) + (\J\otimes\J)\T_{1,1} + \nonumber \\
&\T_{0,2}(\Eye\otimes\J^2) + 2(\Eye\otimes\J)\T_{0,2}(\Eye\otimes\J) + (\J^2\otimes\Eye)\T_{0,2}.
\end{align}
Here, we adopt the convention that all tensor products have the matrices in the $i$ or $i'$ dimension on the left-hand side of the $\otimes$ symbol and $j$ or $j'$ matrices on the right-hand side. In fact, it turns out that the exact number of polynomial terms is 
\beq
N_\text{poly} = \frac{1}{24}\left(24 + 50\omega + 35\omega^2 + 10\omega^3 + \omega^4\right), \label{eq:NPolyTerms}
\eeq
where $\omega \equiv \max(n+m)$ is the highest order polynomial considered. 

\begin{figure}[] 
	\centering 
	\includegraphics[width=.5\textwidth]{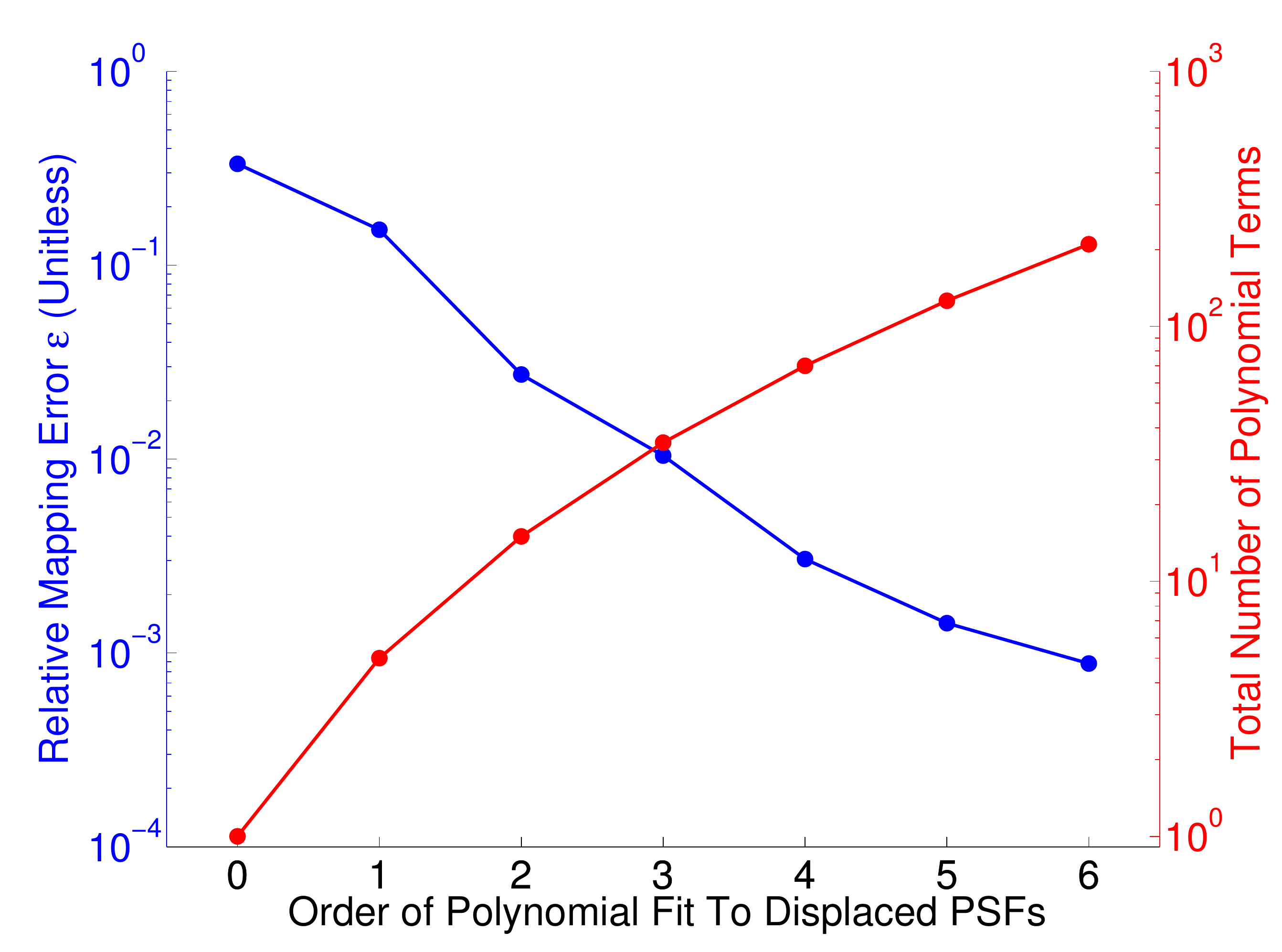}
	\caption{Approximating the translational variation of the point spread function with a low order polynomial can produce fairly small errors at a relatively low accuracy cost. Here we show the accuracy of multiplying a polynomially approximated $\PSF$ with the true sky compares to a direct calculation (using a large PSF truncation radius and no snapshotting). The errors are not negligible and the use of this approximation requires a carful examination of the accuracy requirements of the dirty maps. This technique saves time when the total number of terms in a polynomial/Toeplitz expansion of $\PSF$ is considerably smaller than the number of pixels in a facet. Unfortunately, that number of terms grows quartically with the polynomial order, meaning that very high orders and thus very high accuracy are not computationally useful.}
	\label{fig:PSF_Fitting_Order}
\end{figure}

The good news is that this fitting works pretty well at relatively low order, such as cubic or quartic. In Figure \ref{fig:PSF_Fitting_Order} we calculate the relative error between a dirty map computed by convolving the pixelized ``true" sky with a very accurate $\PSF$ (one computed with a large truncation radius and no snapshotting) and one computed with a polynomial fit to the translationally varying component of $\PSF$.  We find that the method outlined above can faithfully reproduce the dirty map to high precision.

Increasing accuracy, however, comes at a steep cost. While multiplication of $\PSF$ by a vector for a single frequency can be performed in $\BigO{N_\text{facet}N_\text{PSF}}$, multiplication of a polynomially-approximated $\PSF$ takes $\BigO{N_\text{poly}N_\text{PSF}\log N_\text{PSF}}$. Since $N_\text{poly}$ scales with the fourth power of the maximum order, it gets expensive very quickly. Thus the method outlined above is especially useful when $\sim 1\%$ to $0.1\%$ errors are acceptable or when facets are exceptionally big or of exceptionally high resolution. 

It is possible to reduce that cost by attacking the problem with a hybrid approach. We find that the biggest fitting errors come far from the facet center, especially in the brightest side lobes. This makes sense, since it is where the notion of a fixed ``displacement" from the main lobe of the PSF runs up against the limits of the flat-sky approximation. One could use this technique to incorporate the effects of most of $\PSF$, zeroing out the contributions from side lobe displacements. Then we could take the remainder of the $\PSF$ into account by simple matrix multiplication, achieving the same error with many fewer polynomial terms.

With big facets or at high resolution, PSF fitting serves another function. If the computational cost of mapmaking and  power spectrum estimation is dominated by the matrix multiplication $\A^\dagger \N^{-1} \A$ in the calculation of $\PSF$, we can choose to calculate only a representative sample of the entries in $\PSF$ (i.e. only some of the points on the right-hand side of Figure \ref{fig:PSF_Fitting_Demo}). Then we would rely on the fact that the polynomial fit is overdetermined to back out the missing entries.\footnote{It is worth noting that although a large number of terms might be needed to multiply $\PSF$ by a vector, there are not nearly so many free parameters in the fits. The number of free parameters needed to find a best-fit surface like that in Figure \ref{fig:PSF_Fitting_Demo} only scales like the square of the highest polynomial order.}

Whether or not to use the polynomial approximation to the $\PSF$ will depend on the exact telescope configuration and the nature of the mapmaking and power spectrum estimation problems at hand. If we want to try to precisely subtract foregrounds and work deep within the wedge, the polynomial approximation might not be good enough. However, if instead our power spectrum estimation strategy is to focus on isolating the EoR window and projecting out foreground-dominated modes entirely, it is less important that we very precisely understand the effect of the instrument. In that case, it is more likely that the polynomial PSF fitting approach outlined above will be useful. We explore these two approaches in the context of the mapmaking formalism in Appendix \ref{app:projection}.

\subsection{Computational Methods Summary} \label{sec:computationalSummary}

In the previous three sections, we explored three different ways of speeding up either the calculation of $\PSF$ or the multiplication of $\PSF$ by a vector. In Table \ref{tab:summary} we summarize those results. In general, we find that PSF truncation and snapshotting have the most utility for HERA. PSF fitting, in the fiducial scenario we considered, is the least helpful. However, for a telescope with much higher angular resolution than HERA, PSF fitting is likely to be more useful, since multiplication of a vector by $\PSF$ scales quadratically with the number of pixels in the facet.
\begin{table*}[t]
  \centering
\begin{ruledtabular}
  \begin{tabular}{ r  c  c  c }
\textbf{Approximation Parameter} & \textbf{PSF Truncation Radius} & \textbf{Snapshot Time} & \textbf{PSF Fitting Order} \\ \hline 
 \textbf{Improves} & Size of $\PSF$ & Steps in computing $\PSF$ & Multiplying by $\PSF$ \\ 
 \textbf{More exact when...} & ...larger & ...smaller & ...larger \\ 
 \textbf{Cost scaling} & Quadratic & Inverse Linear & Quartic \\ 
 \textbf{Acheives 1\% Error\footnote{Assumes HEALPix $N_\text{side}=256$, 2\,s integrations, and $10^\circ$ diameter facets.} for HERA at} & $5^\circ$ or\footnote{This depends on whether point sources are included, since they are mostly inside the facet in our simulations, depressing the error at small truncation radius.} $15^\circ$ & 10 minutes & $3^\text{rd}$ order \\ 
 \textbf{Speed-up at 1\% error} & $\sim$60 or $\sim$500 & $\sim$300 & $\sim$5 \\ 
  \end{tabular}
\end{ruledtabular}
  \caption{Summary of the techniques we use to approximate the calculation of multiplication by the matrix of point spread functions, $\PSF$, in order to dramatically improve the speed of those operations. Faceting alone makes calculating $\PSF$ faster by a factor of 500 in our fiducial scenario. Combining PSF truncation and snapshotting brings the calculation of $\PSF$ well within the realm of feasibility. The benefit to fitting the PSF with polynomials and Toeplitz matrices is relatively small for our scenario, but it gets much better for higher resolution instruments.}
  \label{tab:summary}
\end{table*}

While these results are specific to HERA, we can draw a few general conclusions. For HERA at 150\,MHz, the first side lobes are about $13^\circ$ from the main lobe of the synthesized beam. At $13^\circ$ from the zenith, the primary beam is down by 20\,dB. In general, it is likely we will only be able to truncate the PSF in regions where the primary beam is small, meaning that a telescope with a broader primary beam will benefit less from cropping in a way that scales quadratically with the PSF truncation radius and therefore also the PSF's full width at half maximum. By contrast, larger primary beams are more slowly varying spatially, meaning that longer snapshots are likely to achieve the same error. If the primary beam is relatively smooth, that benefit scales inverse-linearly with the size of the primary beam.

Though we used 1\% as a somewhat arbitrary point of comparison in Table \ref{tab:summary}, it remains an open question how good our models of the $\PSF$ have to be. The only comprehensive way to answer this question is through a full end-to-end simulation of the signal, noise, and foregrounds all passed through a simulated instrument, a mapmaking code, and then power spectrum estimation. That sort of quantitative answer is outside the scope of this paper. However, it is worthwhile to enumerate the ways in which we need to use $\PSF$ to make maps and estimate power spectra and to examine the accuracy requirements for those tasks. By our count, $\PSF$ appears in six key places in the power spectrum estimation process:
\begin{enumerate}
\item When we calculate $\widehat{\x}$, we need $\PSF$ to define $\D$. However, looking closely at Equation \eqref{eq:QE} shows that $\D$ actually cancels out---the factor of $\D$ in each $\widehat{\x}$ and the two in $\C,_\beta$ are canceled by the two in each $\C^{-1}$. Therefore, it does not matter whether we get $\D$ right or not, as long as we are consistent about what we use for it. This makes sense, $\D$ was supposed to be an arbitrary choice, so as long as it is invertible, there is no way to get it ``wrong" per se.
\item $\PSF$ also appears in our models for the parts of $\mean$ and $\C^{FG}$ corresponding to bright point sources in $\C^{FG}$. Accounting properly for bright point sources has the highest bang for the buck, in the sense that it is relatively straightforward to model both their means and covariances in the dirty map. In Section \ref{sec:skymodel}, we discussed how we could account for bright point sources with well-characterized positions, fluxes, and spectral indices by calculating a column in $\PSF$ that maps the point source to the entire facet in the dirty map. For that calculation, the PSF truncation radius is irrelevant because we account for the brightest sources in a separate part of the PSF independent of the HEALPix grid. Since we calculate only a moderate number of columns of $\PSF$, we do not even have to combine integrations into the snapshot. For bright point sources, it is not much extra effort to get $\PSF$ almost exactly right.
\item By contrast, diffuse emission from confusion-limited and galactic synchrotron emission in $\mean$ and $\C^{FG}$ depends, as we have argued, on knowing how $\PSF$ maps a large part of the true sky onto the facet. It is in this context that approximate versions of $\PSF$ are the most useful, but also where they are potentially the most worrisome. Galactic and confusion-limited foregrounds are still orders of magnitude stronger than the cosmological signal and understanding them precisely is very important. Forming $\mean$ from these foregrounds should be comparatively easy---all we need to do is take our sky model, compute visibilities, and then pass it through our mapmaking routine. We do not even need to calculate the full $\PSF$ matrix. Writing down $\C^{FG}$ is substantially more difficult, since $\C^{FG} = \PSF \C^{FG}_\text{model} \PSF^\trans$. Exactly how well we need to know $\PSF$ in order for $\C^{FG}$ to accurately reflect the foreground uncertainty depends on the specific instrument, the foreground model, and our uncertainty about that model. A quantitative answer requires detailed covariance modeling outside the scope of this work and is therefore left for future investigation.
\item Modeling noise properly is extremely important since inside the EoR window only noise and signal should matter. A slight mismodeling of noise due to an error in the calculation of $\C^N$ could lead to an erroneous detection. If however we perform mapmaking twice from a cross power spectrum of interleaved timesteps, we can eliminate noise bias \cite{X13,PAPER32Limits}. If we do that, it is acceptable (albeit not optimal) to be very conservative in our model of the instrumental noise, effectively increasing the error bars due to noise without biasing our measurement. If we adopt this conservative stance, then we can confidently use an approximate form of $\PSF$ when calculating $\C^N$.
\item Modeling $\C^S$ is mostly important for the calculation of sample variance. In any foreseeable experiment, this is a small contribution to the error. Getting $\C^S$ slightly wrong is unlikely to be the dominant error associated with approximating $\PSF$.
\item The $\C,_\beta$ family of matrices is necessary for telling us how to translate properly weighted dirty maps into power spectra. We  need $\PSF$ to be as accurate as the precision with which we would like to measure the power spectrum. 
\end{enumerate}
In general, the question of exactly how accurately we need to know $\PSF$---and by extension, exactly how well we need to understand our instruments---is an open question for future investigation.

\section{Summary and Future Directions} \label{sec:Summary}

In this work, we showed how to make precise maps with well-understood statistics specifically for 21\,cm power spectrum estimation. We investigated how to connect the framework of optimal mapmaking to that of inverse-variance weighted quadratic power spectrum estimation in order to understand what sort of maps and map statistics we need for power spectrum estimation. We showed that in addition to the dirty map estimator $\widehat{\x}$, we need the matrix of point spread functions, $\PSF$, and the noise covariance matrix which takes a gratifyingly simple form: $\C^N = \PSF \D^\trans$ where $\D$ is an invertible normalization matrix that we can choose to be diagonal. 

This analysis technology will allow us to consistently integrate our best understanding of an instrument with our best models for noise, foregrounds, and the cosmological signal. Not only does this approach help prevent the loss of cosmological information, but it will allow for a precise measurement of the 21\,cm power spectrum and for the confident and robust description of the errors in our estimates. 

In the main part of this work, we focused on the matrix of point spread functions, $\PSF$, which relates the true sky to our dirty maps. We calculated simulated dirty maps and PSFs for HERA, the upcoming Hydrogen Epoch of Reionization Array. While calculating $\PSF$ exactly is computationally prohibitive, we explored three methods for approximating $\PSF$. First, we explored how making maps in facets with truncated PSFs can dramatically reduce the computational cost of calculation $\PSF$ for only a small hit to accuracy. Next we showed how to combine consecutive integrations while controlling for the errors introduced by the process. It turns out that observations many minutes apart can be combined with minimal error. Lastly, we showed how the multiplication of $\PSF$ by a vector---a necessary step for power spectrum estimation---might be sped up by approximating its translational variance as slowly varying. Though the cost scaling of this approximation is steep, we find this technique especially promising when moderate errors are tolerable or for instruments with high angular resolution. 

Just as importantly, all these methods have tunable knobs---they can be made more accurate at the cost of speed or memory. Though our specific, quantitative results are only applicable to HERA, the accuracy trade-offs and the computational scalings we find should be quite general. In that sense, we hope that this work serves as a versatile guide to mapmaking in the context of 21\,cm cosmology.
 
Much work remains to be done to develop a clear and computationally tractable pathway from visibilities all the way to power spectra with rigorous errors and error correlations. Even after connecting this work to an appropriately updated version of the \citet{DillonFast} algorithm, one still needs to assess the effect of our approximations, as well as a number of important data analysis choices, on power spectrum estimates and ultimately on cosmological parameter constraints. Though the errors incurred by each can be made arbitrarily small, it is difficult to say yet what level of approximation is tolerable. This is an open question for future work.

We would like to see a full end-to-end simulation, starting with the 21\,cm signal, passing through the instrument, and ending with power spectra and their statistics. Such a full-scale test could prove the effectiveness of these techniques and clarify exactly what the approximations utilized both in this work and in \citet{DillonFast} do to our measured power spectra. A power spectrum estimation technique that passes such a test with realistic foregrounds and noise will be the one to produce trustworthy cosmological measurements.

\section*{Acknowledgments}
The authors would like to thank James Aguirre, Adam Beardsley, Lu Feng, Bryna Hazelton, Alex Ji, Daniel Jacobs, Alan Levine, Jonathan Pober, Ron Remillard, Michael Valdez, and Ian Sullivan for helpful discussions. We would also like to thank David DeBoer for the rendering of HERA in Figure \ref{fig:HERA-331} and S\'{e}bastien Loisel for the idea behind Equation \eqref{eq:polyToeplitz}. This work was supported by NSF grants AST-0457585, AST-0821321, AST-0804508, AST-1105835, AST-1125558, AST-1129258, AST-1410484, and AST-1411622, a grant from the Mt. Cuba Astronomical Association, the MIT School of Science, the Marble Astrophysics Fund, and by generous donations from Jonathan Rothberg and an anonymous donor.

\appendix

\section{Polarization and Heterogenous Primary Beams} \label{app:PolAndBeams}

In Section \ref{sec:interferometry}, we worked out the relationship between visibilities and the true sky in terms of the matrix $\A$. For the sake of simplicity, we made two assumptions that, in this appendix, we would like to relax.

First, we ignored the effect of polarization. Though the 21\,cm signal is unpolarized, astrophysical foregrounds are generally polarized. And because the primary beams of the two orthogonal polarizations measured by a single element are different, the polarization of sources is important. This is especially important for sources with high rotation measures \cite{MoorePolarization}. Second, we ignored the possibility that not every element has the same primary beam. It is possible that an array is intentionally constructed with multiple kinds of elements. It is also generally true that different elements will behave slightly differently, just due to the variations in their construction. If we are able to measure that variation---which is no small task---we would like to take it into account.

Let us begin with polarization. There are a number of different conventions for expressing polarization \cite{FFTT}, but one relatively straightforward one is to replace $I(\rhat,\nu)$ with a four-element vector $\Eye(\rhat,\nu)$ containing Stokes I, Q, U, and V parameters. Instead of one visibility per baseline and frequency, we now measure four, one for each of the pairs of polarizations of antennas, $xx$, $xy$, $yx$, and $yy$. In this case, $B(\rhat,\nu)$ becomes $\mathbf{B}(\rhat,\nu)$, a $4\times 4$ matrix that describes the response of each type of visibility to each polarization and direction.

Otherwise, not much changes. The sky vector we are estimating gets four times bigger and the number of visibilities also gets four times bigger (though there are simplifications in practice, since $xy$ and $yx$ visibilities are just complex conjugates of one another and can be averaged together to reduce noise). The $\A$ matrix is not fundamentally different. Though it may seem like this makes the problem of computing $\PSF$ 64 times harder, that is fortunately not the case.

Fundamentally, we want to estimate the cosmological signal from our best guess at the Stokes I map. Foregrounds can have I, Q, and U components---astrophysical sources are not circularly polarized. So what we really want is a $\PSF$ matrix that maps I, Q, and U on the true sky, through $xx$ and $yy$ visibilities, to a dirty map of Stokes I. That is only six times more difficult than the calculations outlined above. If we do not want to model our foreground residual as polarized, then $\PSF$ is only twice as complicated as before---we just need to calculate $\mean$ through a more complicated mapmaking procedure involving the expanded definition of $\A$.

The issue with polarization is in many ways similar to the problem of heterogeneous primary beams. After all, the two polarization's dipoles generally have two different primary beams. Since the calculation of $\A^\dagger \N^{-1} \A$ is the computationally limiting step in our method, it is not significantly more difficult to treat multiple kinds of primary beam products $B(\rhat,\nu)$ when calculating $\A$, each row having a potentially different $B(\rhat,\nu)$. This gives us a straightforward way to account for arrays that include multiple types of antenna elements.

Of greater concern is the fact that every element in a real array has a slightly different beam---even if it was designed to be homogenous. For a minimally redundant array, this does not matter. If we know the correct primary beam for every antenna, we can write down $\A$ exactly. For a highly redundant array like HERA, antenna heterogeneity breaks the redundancy of baselines. If we want to include all measured visibilities in our maps, we may need to treat visibilities involving the most discrepant antennas separately. If we had to go further and treat every visibility separately, that would make $\PSF$ two orders of magnitude more difficult to calculate for HERA. If we can measure primary beams for all of our antennas, it would be worthwhile to simulate the error associated with the approximation that they are all the same. This is left to future work. Fortunately, it is theoretically possible to take into account slight variations between elements in the framework we have outlined.

\section{A Foreground Avoidance Approach to Power Spectrum Estimation} \label{app:projection}

The power spectrum estimation method we outlined in Section \ref{sec:powerspectra} is a promising way to enlarge the EoR window and gain the additional sensitivity forecasted by \cite{PoberNextGen}. However, it is not the simplest approach. Instead of directly modeling foregrounds, we could choose to simply throw out all the modes that we believe to be foreground contaminated. The foreground avoidance approach was pioneered by \cite{AaronDelay} and used to produce the best current limits on the 21\,cm power spectrum by \cite{PAPER32Limits} and \cite{DannyMultiRedshift}. This choice should be more robust to foreground mismodeling than subtraction, since we are merely trying to isolate foreground free regions of Fourier space from the effects of regions we have given up on. Where exactly we draw the line between wedge and window is a question that deserves further investigation with both simulations and real data. 

One might ask why foreground avoidance estimators are interesting when the whole point of making maps like ours was to compress the data in a space where foregrounds were most naturally subtracted. There are a few reasons. First, foreground avoidance is simpler than foreground subtraction. If we are going to try to subtract foregrounds, it is worthwhile to first perform the simpler, more robust procedure so we have a baseline for comparison. Second, even if we are only interested in mitigating the effect of foregrounds by avoiding them, this method gives a proper accounting for $\C^N$, $\C^S$, and $\C,_\alpha$, without making any of the approximations previously relied upon about there being no correlations between $uv$ cells or that uniform weighted maps have no PSF. Third, the technique is fairly directly comparable to that of \cite{AaronDelay} without the additional assumption that delay modes for a given visibility map neatly to band powers or the computational challenges of \cite{EoRWindow1,EoRWindow2}. And finally, we may also want to implement a hybrid approach, similar in spirit to \cite{PAPER32Limits}, where we project out modes deep into the wedge but try to subtract foregrounds nearer the edge of the wedge.\footnote{This is similar in spirit to what WMAP did \cite{WMAP1PowerSpectrum}. They first masked out the galaxy and the brightest point sources, then they performed foreground subtraction in the map and foreground residual bias subtraction in the angular power spectrum. For us, the major difference is that the we do both } 

Therefore, it is worthwhile to write down the general framework for foreground avoidance in the context of optimal mapmaking. The idea is relatively simple. Let's define a new dirty map estimator, $\widehat{\x}'$, defined as
\beq
\widehat{\x}' \equiv \Proj \widehat{\x}
\eeq
where $\Proj$ is a projection matrix that has eigenvalues of 0 or 1 only.  As with all projection matrices, $\Proj = \Proj^\trans = \Proj^2$. The matrix $\Proj$ Fourier transforms the data cube, sets all modes outside the EoR window to zero, and Fourier transforms back. It also means that we need to replace $\C$ with $\C'$ where
\beq
\C' = \Proj \C \Proj.
\eeq 
 
By construction, the projection eliminates the foregrounds in $\mu$, meaning that
\begin{equation}
\Proj \langle \widehat{\x} \rangle = \Proj \mu \approx 0.
\end{equation}
Likewise, the part of the covariance associated with the foregrounds should also go to zero. Hence,
\beq
\Proj \C^{FG} \Proj \approx 0, 
\eeq
which means that 
\beq
\C' = \Proj \left[ \C^S + \C^N \right] \Proj.
\eeq
This also changes $\C_{,\alpha}$ which now takes the form
\beq
\C_{,\alpha}' = \Proj \C_{,\alpha} \Proj = \Proj \PSF \Q_\alpha \PSF^\trans \Proj.
\eeq

Of course, the new covariance has many zero eigenvalues, which means that it is not invertible.  That is not a problem since we can replace $(\C')^{-1}$ by its ``pseudoinverse" \cite{Maxgalaxysurvey1}, defined as
\beq
(\C')^{-1}_\text{psuedo} = \Proj \left[ \Proj \C' \Proj + \gamma(\Eye - \Proj)\right]^{-1}\Proj
\eeq
where $\gamma$ can be any (numerically reasonable) nonzero number without changing the result.  The pseudoinverse reflects the idea that we want to completely throw out any power in possibly foreground-contaminated modes but also that we want to express infinite uncertainty in the modes---in other words, to give them no weight. This will accurately account for the fact that we have no information about these modes.

Putting all that together, our new quadratic estimator $\widehat{\p}$ is
\begin{align}
\widehat{p}_\alpha =& \frac{1}{2} M_{\alpha \beta} \widehat{\x}^\trans    (\C')^{-1}_\text{psuedo}  \PSF \Q_\beta \PSF^\trans  
(\C')^{-1}_\text{psuedo}  \widehat{\x} - b_\alpha,
\end{align}
where we have used the fact that $\Proj^2 = \Proj$. The estimator is not lossless, but it can still be unbiased in the region of Fourier space not projected out and have rigorously defined and calculable error properties.

\bibliography{Precision_Mapmaking}  

\end{document}